\documentclass[11pt,a4paper]{article}
\usepackage{jheppub}
\usepackage{amsmath,amsfonts,amssymb,amsthm,graphics,graphicx,epsfig,times,bbm,verbatim}
\def\beqn{\begin{eqnarray}}
\def\eeqn{\end{eqnarray}}
\def\beqs{\begin{subequations}}
\def\eeqs{\end{subequations}}

\newcommand\be{\begin{equation} }
\newcommand\ee{\end{equation}}
\newcommand\IC{{\mathbb{C}}}
\newcommand\IH{{\mathbb{H}}}
\newcommand\IQ{{\mathbb{Q}}}

\newcommand\IZ{{\mathbb{Z}}}
\newcommand\CM{{\cal M}}

\def\CM{{\cal M}}
\def\CP{{\cal P}}

\title{Hecke Relations in Rational Conformal Field Theory}

\author{Jeffrey A. Harvey}
\author{and Yuxiao Wu}
\affiliation{Enrico Fermi Institute and Department of Physics University of Chicago \\
933 E 56th St.,  Chicago IL 60637}

\emailAdd{j-harvey@uchicago.edu}
\emailAdd{yuxiaowu@uchicago.edu }

\abstract{We define Hecke operators on vector-valued modular forms of the type that appear as characters of rational conformal field theories (RCFTs).
These operators extend the previously studied Galois symmetry of the modular representation and fusion algebra of RCFTs to a relation between RCFT characters. We apply our results to derive
a number of relations between characters of known RCFTs with different central charges and also explore the relation between Hecke operators and RCFT characters as solutions to modular linear differential equations.
We show that Hecke operators can be used to construct an infinite set of possible characters for RCFTs with two independent characters and increasing central charge.  These characters have multiplicity one for the vacuum representation,
positive integer coefficients in their $q$ expansions, and are associated to a two-dimensional representation of the modular group which leads to non-negative integer fusion coefficients as determined
by the Verlinde formula.
\vskip 0.1in
\today}

\begin{document}

\maketitle

\section{Introduction}\label{sec:Intro}

Two-dimensional rational conformal field theory (RCFT) plays a central role in many different areas of physics and mathematics ranging from critical phenomenon, boundary modes in topological insulators and the world-sheet description of classical string theory backgrounds in physics to the modularity theorem and the study of sphere packing in mathematics, to name two relatively recent examples.  The characters of RCFTs are modular forms of a type that will be described later. Hecke operators and their eigenfunctions play a crucial role in the mathematical theory of modular forms, thus it is natural to expect that they should play a role in RCFT as well. Hecke
operators have in fact appeared in physics, an important example is the work \cite{Dijkgraaf:1996xw} where they were used in the computation of the elliptic genus of the symmetric product
of K{\"a}hler manifolds. Here we find a new application: they can be used to relate characters of  RCFTs with different central charges and to construct infinite classes of modular functions which have all the properties required of characters of consistent RCFTs.

There is a fundamental
difference between the modular forms appearing in RCFT and those appearing in many areas of mathematics.  To explain this first recall that $SL(2,\IZ)$ acts on $\tau$ in the upper half plane $\IH$
by
\be
\tau \rightarrow \gamma \tau = \frac{a \tau +b}{c \tau +d}
\ee
with
\be
\gamma = \begin{pmatrix} a & b \\ c & d \end{pmatrix} \in SL(2,\IZ) \, .
\ee
A modular form of weight $k$ for the modular group $\Gamma=SL(2,\IZ)$
is a function $f: \IH \rightarrow \IC$  which is holomorphic  (except for a possible pole as $\tau \rightarrow i \infty$) in $\IH$ and which obeys the transformation law
\be \label{ftransf}
f( \gamma \tau) = (c \tau+d)^k f(\tau) \, .
\ee
Invariance under $\tau \rightarrow \tau+1$ implies that modular forms have an expansion in $q=e^{2 \pi i \tau}$ of the form
\be
f(\tau)= \sum_n a(n) q^n \, .
\ee
Modular forms can be distinguished by their behavior as $q \rightarrow 0$ or $\tau \rightarrow i \infty$ (the cusp at infinity). If $f$ is holomorphic away from the cusp
at infinity but diverges at the cusp with a finite number of $a(n) \ne 0$ for $n<0$ then $f$ is said to be weakly holomorphic. If $f$ is holomorphic everywhere including at the cusp,
implying that $a(n)=0$ for $n<0$ then $f$ is usually just called a modular form. If $f$ vanishes at the cusp at infinity so that $a(n)=0$ for $n \le 0$ then $f$ is called a
cusp form.  We will follow standard usage by referring to functions obeying eqn.(\ref{ftransf}) with weight $k=0$ as modular functions rather than modular forms.

A famous example of a weight $12$ cusp form appears as the inverse of the oscillator partition function of bosonic string theory,
\be
\Delta(\tau) = (\eta(\tau))^{24}= q \prod_{n=1}^\infty(1-q^n)^{24} \, .
\ee
Hecke operators will be described in more detail later, for the moment we define the operator $T_m$ for $m \in \IZ$ through its action on the coefficients of the $q$ expansion
of a modular form. If $f= \sum_n a(n) q^n$ is a weight $k$ modular form then we define \footnote{Later we will find it convenient to choose a different  normalization for $T_m$
than that given here.}
\be
(T_m f)(\tau) = \sum_n a^{(m)}(n) q^n
\ee
with
\be
a^{(m)}(n) = \sum_{s>0, s|\,\text{gcd}(m,n)} s^{k-1} a(mn/s^2) \, .
\ee

It can be shown that Hecke operators map cusp forms to cusp forms of the same weight, and since the weight $12$ cusp form is the unique form of this weight, it must be an eigenfunction
of all the $T_m$. In fact one has
\be
T_m \Delta = c(m) \Delta
\ee
where $c(m)$ is the coefficient of $q^m$ in the $q$ expansion of $\Delta$. It is also clear that weakly holomorphic modular forms and functions cannot be eigenfunctions of Hecke operators
since if $n<0$ is the power of $q$ in the most singular term in a weakly holomorphic modular form or function $f$ then $T_m f$ will have most singular term $q^{mn}$.  This applies in particular to characters of conformal field theories. In this case the individual characters are modular functions for the principal congruence subgroup $\Gamma(N)$ for a finite $N$ defined below, and the components transform under a finite-dimensional representation $\rho$ of the full modular group $SL(2, \IZ)$ with a kernel containing $\Gamma(N)$. At least one of the characters is always weakly holomorphic with leading term
$q^{-c_{\rm eff}/24}$ with $c_{\rm eff}>0$ and although the precise form of the Hecke operators is different, the behavior  is similar in that the resulting image under the action of the Hecke
operator has more singular behavior as $q \rightarrow 0$.  Therefore characters of rational conformal field theories cannot be eigenfunctions of Hecke operators.
However this does not rule out the possibility that Hecke operators might relate the characters of RCFTs with different values of $c_{\rm eff}$ and in fact we will find that suitably defined Hecke operators do in fact relate characters of many RCFTs. This work can be seen as a small step towards extending some of the mathematical machinery used to study cusp forms to the world of weakly holomorphic weight zero modular functions and the physics of two-dimensional conformal field theory. 

Our first main result consists of a purely mathematical statement about a particular class of Hecke operators that act on vector-valued
weakly holomorphic modular functions of a particular type. 
We define a level $N$ vector-valued modular function $f$ with respect to a representation $\rho$ \footnote{This definition can obviously be extended to non-zero weight, but for simplicity we restrict to weight zero since this is the
case arising in RCFT.} to be a set $f_i(\tau)$, $i=1, \cdots n$ 
of weakly holomorphic modular functions for $\Gamma(N)$ which transform under $SL(2,\IZ)$ as
\be
f_i(\gamma \tau) = \sum_j \rho(\gamma)_{ij}  f_j(\tau)
\ee
where $\rho$ is a $n$-dimensional unitary representation
\be
\rho: SL(2, \IZ) \rightarrow GL(V(\rho))
\ee
with $V$ a complex vector space of finite dimension $n$ and where $N$ is the finite order of $\rho(T)$ and $\Gamma(N)$ is in the kernel of $\rho$. Here
\be
T= \begin{pmatrix} 1 & 1 \\ 0 & 1 \end{pmatrix}, \qquad S= \begin{pmatrix} 0 & -1 \\ 1 & 0 \end{pmatrix}
\ee
are the generators of $SL(2, \IZ)$.  All characters of RCFTs are level $N$ modular functions for some finite-dimensional representation $\rho$ by a theorem of Bantay \cite{Bantay:2001ni} 
but the converse is not true. 

If $f$ is a level $N$ modular function with respect to the representation $\rho$, then for each $p$ such that $\text{gcd}(p,N)=1$ there are Hecke operators ${\mathsf T}_p$ such that 
$({\mathsf T}_p f)_i$ are the components of a level $N$ modular function with respect to the representation 
$\rho^{(p)}$ defined by its values on the generators
of $SL(2,\IZ)$,
\be \label{newrep}
\rho^{(p)}(T)  = \rho(T^{\bar p}), \qquad \rho^{(p)}(S)= \rho( \sigma_p S) \, ,
\ee
where $\bar p$ is the multiplicative inverse of $p$ mod $N$ and $\sigma_p$ is the preimage of
\be
\begin{pmatrix} \bar p & 0 \\ 0 & p \end{pmatrix}
\ee
under the mod $N$ map from $SL(2, \IZ)$ to $SL(2, \IZ/N \IZ)$. The mod $N$ map takes an element of $SL(2, \IZ)$ and reduces every element of the matrix mod $N$ to produce
an element of $SL(2, \IZ/N \IZ)$. Note that $\Gamma(N)$ is the kernel of the mod $N$ map and hence a normal subgroup of $SL(2, \IZ)$. 
Note that $\sigma_p$ is not unique, but since two choices differ by the action of
$\Gamma(N)$ which is in the kernel of $\rho$, the representation $\rho^{(p)}$ is uniquely determined by any choice of $\sigma_p$. The Hecke operators
${\mathsf T}_p$ are defined in section \ref{sec:HeckeRCFT} where a number of relations they obey are also discussed. Note that the order of $\rho(T^{\bar p})$ is the same as the order
of $\rho(T)$. Thus neither the value of $N$ nor the dimension of the representation changes under the action of ${\mathsf T}_p$. 

Our second main result is that there are many sets of RCFT characters that are related by the action of the Hecke operators ${\mathsf T}_p$. A number of examples are discussed in Section \ref{sec:Applications}.  Our third result is that the Hecke operators $\mathsf{T}_p$ for a certain infinite set of values of $p$ can be used to construct infinite sets of level $N$ modular functions associated to a finite set of representations $\rho^{(p)}$ that satisfy all the known constraints for characters of a RCFT. In particular they have degeneracy one for the vacuum representation, have all non-negative integer coefficients in their $q$ expansions, and the fusion coefficients determined from the Verlinde formula are all non-negative. The simplest example is an infinite set with two independent characters which arise as Hecke images of the Yang-Lee model. 

The outline of this paper is as follow. In section \ref{sec:RCFT} we review the general structure of RCFT and discuss two results of particular relevance to us, namely the appearance of a Galois symmetry  which relates representations of $SL(2,\IZ)$ appearing in RCFTs and the fact that characters
of a RCFT are solutions to Modular Linear Differential Equations (MLDE) and we discuss the role played by the number of zeros of the Wronskian in the classification of MLDE.  In Section \ref{sec:ScalarHecke} we review the most basic description of Hecke operators for (scalar) modular forms for the full modular group $\Gamma=SL(2, \IZ)$. We then discuss the extension to Hecke operators for the principal congruence subgroup $\Gamma(N)$ and discuss the difficulties in extending these results to Hecke operators for vector-valued modular forms. 
Section \ref{sec:HeckeRCFT} contains our first main result, showing that there is a specific extension of Hecke operators to level $N$ modular forms with respect to a representation $\rho$. We explain the relation of these Hecke operators to the previously observed Galois symmetry and also discuss the interplay between these Hecke operators and MLDE. We then turn in Section \ref{sec:Applications} to a number of examples of RCFT that have characters related by Hecke operators, focusing for simplicity on RCFT with two or three independent characters. We find a number of relations between minimal models, WZW models and other RCFT that appear in the literature.  Finally in Section \ref{sec:Conclusions} we conclude and offer some thoughts on future avenues of research that are suggested by our results.

\section{RCFT}\label{sec:RCFT}

We first provide a brief summary of RCFT and set up our notation. For more in depth reviews see \cite{Moore:1989vd, Fuchs:2009iz, Huang:2013jza}. 

In a two-dimensional rational conformal field theory there are chiral algebras ${\cal A}$, $\overline{ \cal A}$ such that the Hilbert space
decomposes into a finite number of irreducible representations $V_i, \overline{V_{\bar i}}$ of the chiral algebras and has the form
\be
{\cal H} = \bigoplus_{i \in {\cal I}, \bar i \in \bar {\cal I}} {\cal N}_{i, \bar i} V_i \otimes \overline{V_{\bar i}}
\ee
with ${\cal N}_{i \bar i} \in \IZ_{\ge 0}$. 
The partition function is expressed in terms of characters
\be
\chi_i(\tau) = {\rm Tr}_{V_i} q^{L_0-c/24}, \qquad q=e^{2 \pi i \tau} \, 
\ee
and their anti-holomorphic counterparts
as
\be
Z = \sum_{i \in {\cal I},  \bar i \in \bar {\cal I}} {\cal N}_{i, \bar i} \chi_i(\tau) \overline{ \chi_{\bar i}(\tau) } \, .
\ee
Here we have introduced finite index sets ${\cal I}$, $\bar {\cal I}$ that we use to label the irreducible representations of ${\cal A}$ and $\overline{\cal A}$ as well as the
associated characters. The central charge $c$ is positive in a unitary RCFT and for unitary RCFT the character with most singular behavior as $q \rightarrow 0$ is the
vacuum character with leading behavior $\chi_0 \sim q^{-c/24}$. In non-unitary RCFT $c$ can be negative as can the conformal weights $h_i$ of primary operators. In this
case we define $c_{\rm eff}$ so that the leading singular behavior of the most singular character is $q^{-c_{\rm eff}/24}$. 

We will often use Virasoro minimal models as examples of RCFTs. These are determined by a pair of coprime integers $(p,q)$ with $p >1, q>1$ and
have central charge
\be
c(p,q)= 1 - 6 \frac{(p-q)^2}{pq} \, .
\ee
These models are unitary provided that $|p-q|=1$. Both unitary and non-unitary minimal models will enter into our considerations.
In the following we denote the minimal model labelled by $(p,q)$ by $\mathsf{M}(p,q)$.

\subsection{Modular Properties of Characters}
\label{subsec:modprop}

The modular properties of the characters $\chi_i(\tau)$ will play a central role in what follows. Under  modular transformations 
by $\gamma \in SL(2, \IZ)$ they transform as
\be
\chi_i(\gamma \tau) = \sum_j \rho(\gamma)_{i j} \chi_j(\tau) \, .
\ee
Here
\be
\rho: SL(2, \IZ) \rightarrow GL(V)
\ee
is a unitary, finite-dimensional representation of the modular group with $V$ a complex vector space and $\rho(\gamma)_{ab}$ are the matrix elements of $\rho$ with respect to a basis of $V$. The partition function is required to be modular invariant, thus ${\cal N}$ is an intertwiner between $\rho$ and $\bar \rho$. 
The representation $\rho$ is determined by
its value on the generators $S: \tau \rightarrow -1/\tau$ and $T: \tau \rightarrow \tau+1$ of $SL(2, \IZ)$ and must satisfy
\be
(\rho(S))^2 = (\rho(S) \rho(T))^3 = C
\ee
where $C$ is the conjugation matrix and is an involution, $C^2=1$.

 In a RCFT the central charge $c$ and
conformal weights $h_i$ of primary operators are rational \cite{Anderson:1987ge} and as a result $\rho(T)$ will have finite order which we denote by $N$. One can choose
a basis in which $\rho(T)$ is diagonal and we will do so here. The diagonal elements of $\rho(T)$ are then $N$th roots of unity. 

A theorem of Bantay \cite{Bantay:2001ni} states that the individual characters $\chi_i(\tau)$ are weight zero modular functions for the principal congruence subgroup
$\Gamma(N)$ defined as
\be
\Gamma(N) =  \left \{ \begin{pmatrix} a & b \\ c & d \end{pmatrix} \in SL_2(\IZ) | ~ a \equiv d \equiv 1 ~{\rm mod}~N, ~b \equiv c \equiv0 ~{\rm mod}~N \right \} \, .
\ee
Thus we have $\chi_i( \gamma \tau) = \chi_i(\tau)$ for $\gamma \in \Gamma(N)$. It will be useful later to note that upon reduction mod $N$ of each
element of a matrix $\gamma \in SL(2, \IZ)$ we obtain an element of $SL(2, \IZ/N \IZ)$. The kernel of this mod $N$ map is $\Gamma(N)$ which is thus
a normal subgroup of $SL(2,\IZ)$. 

While all characters of RCFT $\chi_i$ are individually modular forms for $\Gamma(N)$ and transform according to a representation $\rho$ of 
$SL(2, \IZ)$,  the converse is not true. One constraint that any set of modular forms of this type must obey in order to be characters of a RCFT arises from the Verlinde formula \cite{Verlinde:1988sn}. The fusion coefficients ${N_{ij}}^k$ which govern the fusion of primary operators $\phi_i$
\be
\phi_i \times \phi_j = {N_{ij}}^k \phi_k \, 
\ee
are determined in terms of the representation $\rho(S)$ via
\be
{N_{ij}}^k = \sum_m \frac{\rho(S)_{im} \rho(S)_{jm} \rho(S^{-1})_{km}}{\rho(S)_{0m}}
\ee
where the index $0$ labels the vacuum representation. In an RCFT the fusion coefficients must be non-negative integers.

\subsection{Galois Symmetry of RCFT Characters}
\label{subsec:Galois}

The idea of a Galois action on the modular representation $\rho$ of a RCFT originated with work of De Boer and Goeree \cite{DeBoer:1990em}. This Galois action has implications not just for RCFT but also in the study of topological phases of matter, see for example \cite{Freedman}. The basic idea is to construct a field $K$ which
is an extension of the rationals $\IQ$ obtained by adjoining all of the matrix elements of $\rho(\gamma)$ for $\gamma \in SL(2, \IZ)$ to $\IQ$. One then uses Verlinde's formula to show
that $K$ is a finite Abelian extension of $\IQ$ \cite{DeBoer:1990em} . A theorem of Kronecker and Weber states that any such $K$ is a subfield of some cyclotomic field, that is an extension of $\IQ$ by a primitive $m$th root
of unity. Writing $\xi_m= e^{2 \pi i/m}$ we have $K \subseteq \IQ[\xi_m]$. In algebraic number theory the smallest $m$ such that a finite Abelian extension $K \subseteq \IQ[\xi_m]$ is known as
the conductor of $K$ and following \cite{Bantay:2001ni} we will also refer to it is as the conductor of a RCFT that gives rise to $K$ by the procedure just described. One of the results of \cite{Bantay:2001ni} states that the conductor of a RCFT is equal to the order of $\rho(T)$. This had been conjectured earlier in \cite{Coste:1999yc, Bauer}. In what follows we use $N$ to denote the order of $\rho(T)$ and
hence the conductor of the underlying RCFT. 

For a given RCFT the Galois group ${\rm Gal}(K/\IQ)$ is homomorphic to ${\rm Gal}(\IQ[\xi_N])$ and the latter is isomorphic to $(\IZ/N \IZ)^\times$, the group of multiplicative units in
$\IZ/N\IZ$, or equivalently, the set of residue classes  $\ell$ in $\IZ/N \IZ$ with $\text{gcd}(\ell,N)=1$. Each element of $(\IZ/N \IZ)^\times$ gives rise to a Frobenius element $f_{N,\ell}$ that leaves
$\IQ$ fixed and maps $\xi_N$ to $\xi_N^\ell$.  In \cite{Coste:1993af} it was shown that these Frobenius elements act in a simple way on the representation matrices $\rho(S)$, $\rho(T)$.
Since $\rho(T)$ is diagonal the action of $f_{N,\ell}$ is simply
\be\label{rho_T_Galois}
f_{N,\ell}: \rho(T) \rightarrow \rho(T)^\ell \, .
\ee
The action on $\rho(S)$ has the general form
\be\label{rho_S_Galois}
f_{N,\ell}: \rho(S)_{i, j} \rightarrow \varepsilon_{\ell}(i) \rho(S)_{\pi_{\ell}(i), j}=\varepsilon_{\ell}(j)  \rho(S)_{i,\pi_{\ell}(j)}
\ee
where $\pi_\ell$ is some permutation of the elements of ${\cal I}$ and $\varepsilon_{\ell} $ is a map from ${\cal I}$ to $\{+1, -1 \}$. The equality above follows from the fact that $ \rho(S)$ is a symmetric matrix.
We can write this transformation in matrix form by defining the monomial matrix $G_{\ell}$:
\begin{align}\label{def_G}
(G_{\ell})_{i,j} =  \varepsilon_{\ell}(i) \delta_{\pi_{\ell}(i),  j} \, ,
\end{align}
and writing the Galois action on $ \rho(S)$ as \footnote{Our convention for the matrix $G_{\ell}$ is the same as that in \cite{Coste:1999yc}.}
\begin{align}
f_{N,\ell} \big(\rho(S) \big)= \rho(S) \, G_{\ell}^{-1}=G_{\ell} \, \rho(S) .
\end{align}
In what follows we will show that this Galois action
on the representation $\rho$ is the same as an action resulting from Hecke operators on the characters of the RCFT, at least for several classes of RCFT.

\subsection{RCFT Characters and MLDE}

It is known that the characters of a RCFT satisfy modular linear differential equations (MLDE). This was used in \cite{Anderson:1987ge} to show
that the central charge $c$ and conformal weights $h_i$  of primary fields in a RCFT are rational numbers. There are two points of view one can take. Given a specific RCFT, the existence of null states in representations of the chiral algebra can be used to derive MLDE obeyed by the characters
of the theory. This point of view was taken in  \cite{Eguchi} where it was used to derive  MLDE for the Virasoro characters of minimal models. The other point of view is to use MLDE as a method for exploring the space of RCFT characters, the idea being that solutions of an $n$th order MLDE with positive integers coefficients of their $q$ expansion will necessarily form a $n$-dimensional representation of the modular group and are thus candidate characters of a RCFT. This point of view was taken in  \cite{Mathur:1988na, Mathur:1988gt, naculich} where it was used to classify RCFT with two independent characters and in \cite{Hampapura:2016mmz} where it was used to classify possible RCFT characters without dimension one operators. Related work in the mathematical literature includes \cite{kanekozagier, Milas:2004, Mason, kaneko, franc, mason2d, francmason, arike, arikeII}. 

Acting on a modular form of weight $k$  the Serre derivative ${\cal D}_k= d/d \tau - \frac{1}{6} i \pi k E_2$ (with $E_2$ the quasimodular Eisenstein series of weight $2$) produces a modular form of weight $k+2$. Therefore acting on weight zero modular functions the iterated derivative
\be
{\cal D}^n =    {\cal D}_{2n-2} {\cal D}_{2n-4} \cdots  {\cal D}_2 {\cal D}_0
\ee
produces a modular form of weight $2n$. A general $n$th order MLDE then takes the form
\be\label{MLDE}
{\cal D}^n f + \sum_{k=0}^{n-1} \phi_k(\tau) {\cal D}^k f=0
\ee
with the $\phi_k(\tau)$ modular forms of weight $2(n-k)$. 
The coefficients $\phi_k(\tau)$ can
be expressed in terms of the Wronskian determinants constructed out of the $n$ linearly independent solutions $f_1,\cdots ,f_n$ of the MLDE as
\begin{align}
\phi_k = (-1)^{n-k} W_k/W \, ,
\end{align}
where
\be
W_k =  \begin{vmatrix} f_1 & f_2 & \cdots & f_n \\
                                                   {\cal D} f_1 & {\cal D} f_2 & \cdots & {\cal D} f_n \\
                                                   \vdots & \vdots &      & \vdots \\
                                                   {\cal D}^{k-1} f_1 & {\cal D}^{k-1} f_2 & \cdots & {\cal D}^{k-1} f_n \\
                                                    {\cal D}^{k+1} f_1 & {\cal D}^{k+1} f_2 & \cdots  & {\cal D}^{k+1} f_n \\
                                                    \vdots & \vdots &      & \vdots \\
                                                     {\cal D}^{n} f_1 & {\cal D}^{n} f_2 & \cdots  & {\cal D}^{n} f_n
                                                     \end{vmatrix} \, ,
 \ee
and $W = W_n$ which we will call the modular Wronskian.       

In RCFT the characters  are known to be holomorphic in the upper half plane except possibly at the cusp at infinity. Thus the Wronskians $W_k$ cannot have poles
except at infinity. However the coefficient functions $\phi_k$ may have poles coming from the
zeros of $W$. It is useful to classify solutions by the ``number of zeros" of $W$ as has been done in earlier literature \cite{Mathur:1988na, naculich,Hampapura:2015cea}.  

If $f$ is a meromorphic function in the upper half plane $\mathbb{H}$ and if $p \in \mathbb{H}$ is a point at which $f$ has a pole or zero then
the integer $n$ such that $f/(\tau-p)^n$ is holomorphic and non-zero at $p$ is called the order of $f$ at $p$ and denoted by ${\rm ord}_p(f)$. 
Now the modular Wronskian $W$ is holomorphic except perhaps at infinity. Define ${\rm ord}_\infty(W)$ be the leading coefficient $\alpha$ in the $q$ expansion
$W=  q^\alpha \sum_{n=0}^\infty a(n) q^n$ \footnote{$W$ transforms under the modular group with a multiplier system given by $\det \rho(\gamma)$ for $\gamma \in SL(2,\IZ)$.}. Then following the notation of \cite{Mathur:1988gt} we define 
\be
\ell(W) = 6 \left(  \frac{1}{2} {\rm ord}_i(W) + \frac{1}{3} {\rm ord}_\omega (W) + \sum'_{p \in {\cal F}}{\rm ord}_p(W) \right)
\ee
where $i=\sqrt{-1}$, $\omega= e^{2 \pi i/3}$ are the orbifold points of orders $2$ and $3$ (fixed by $S$ and $ST$) and $\sum'$ indicates that
the sum runs over all points not equal to $i, \omega$ or $i \infty$. A standard theorem in the theory of modular forms resulting from integrating $d W/W$ around the boundary
of the fundamental domain of the modular group as in \cite{serre} then states that
\be
{\rm ord}_\infty (W) + \frac{\ell(W)}{6} = \frac{k}{12}
\ee
where $k$ is the weight of $W$. 

For an $n$th order MLDE the modular Wronskian $W$ is a modular form of weight $n(n-1)$ and
if the leading behaviors of the
the solutions $f_i$ are $q^{\alpha_i}$ then this condition states that
\be \label{eqn:rroch}
\sum_i \alpha_i +  \frac{\ell(W)}{6}= \frac{n(n-1)}{12}.
\ee

We first consider the $n=2$ MLDE with $\ell=0$.  Then $W$ has no zeros and thus $\phi_0$, $\phi_1$ have no poles.  Therefore $\phi_0$ must be a weight two modular form holomorphic in all of
$\mathbb{H}$ and similarly for $\phi_1$ of weight four. Since there are no modular forms
of weight two and the Eisenstein series $E_4$ is the unique modular form of weight four holomorphic in $\mathbb{H}$, the general equation with $n=2$ and $\ell=0$
involves one free parameter $\mu$ and  takes the form
\be
D^2 f - \frac{1}{6} E_2 D f - \frac{\mu}{4} E_4 f=0
\ee
in terms of
\be
D=\frac{1}{2 \pi i} \frac{d}{d \tau} = q \frac{d}{dq} \, .
\ee

Two solutions can be found in Frobenius form
\be
f^\pm= q^{\alpha^\pm} \sum_{n=0}^\infty f^\pm_n q^n
\ee
with $\alpha_\pm= (1 \pm x)/12$ where $x$ is the positive square root of $1+36 \mu$.  Choices of $\mu$ that leads to solutions with integer $q$ expansions and which have a candidate vacuum character with leading coefficient one were classified in \cite{Mathur:1988na}. See also \cite{Mathur:1988gt, naculich}. 
Excluding the degenerate case of affine $E_8$ at level one which has a single character, nine solutions were found in \cite{Mathur:1988na} which we label in order of increasing $c_{\rm eff}$ as
$X=\{YL, A_1, A_2, G_2, D_4, F_4, E_6, E_7, E_{7 \frac{1}{2}} \}$. 
The first solution consists of the two characters of a non-unitary RCFT  which is the minimal model $\mathsf{M}(2,5)$ and is also known as the Yang-Lee model with $c=-22/5$ and $c_{\rm eff}=2/5$. 
The last solution consists of the 
characters of a model sometimes denoted by $E_{7 \frac{1}{2}}$. This latter model is not an RCFT because the fusion rules computed from the Verlinde formula are not positive integers. It was recently given an interpretation as an Intermediate Vertex Subalgebra in \cite{kawasetsu} following
earlier work on $E_{7 \frac{1}{2}}$ as an Intermediate Lie Algebra \cite{sextonion}. 
The remaining seven solutions are
characters of affine Lie algebras at level $k=1$ with two independent characters.  These seven affine Lie algebras contain as subalgebras the Lie algebras of the Lie groups  $A_1, A_2, G_2, D_4, F_4, E_6$ and $E_7$ \footnote{Note that the $A_2, D_4, E_6$ cases have more than two inequivalent simple
modules, but only two independent characters.}.  These along with $E_8$ form what is known as the Deligne exceptional series of Lie groups and share a number of remarkable properties \cite{deligne}. 
Further information on these models can be found in the tables in \cite{Mathur:1988na, naculich}. 

We note two interesting connections of these two character theories to number theory. First, it is well known that the characters of the Yang-Lee
model are given by
\begin{align} \label{ylone}
\chi^{YL}_0 (q) &= q^{-1/60} G(q) \\
\chi^{YL}_{1/5}(q) &= q^{11/60} H(q) \nonumber
\end{align}
where the functions $G(q)$, $H(q)$ obey the Rogers-Ramanujan identities
\begin{align} \label{yltwo}
G(q) &= \sum_{n=0}^\infty \frac{q^{n^2}}{(1-q)(1-q^2) \cdots (1-q^n)} = \prod_{n=0}^\infty \frac{1}{(1-q^{5n+1})(1-q^{5n+4})} \\
H(q) &= \sum_{n=0}^\infty \frac{q^{n^2+n}}{(1-q)(1-q^2) \cdots (1-q^n)} = \prod_{n=0}^\infty \frac{1}{(1-q^{5n+2})(1-q^{5n+3})} \nonumber
\end{align}
and the ratio of characters $R(q)= \chi^{YL}_{1/5}(q)/\chi^{YL}_0(q)$ is given by the continued fraction
\be
R(q) = \frac{q^{1/5}}{1+\dfrac{q}{1+\dfrac{q^2}{1+\dfrac{q^3}{1+ \ddots}}}} \, .
\ee
This is the starting point for many striking identities developed by Ramanujan and others. For a brief overview see
\cite{berndt}. Second, if we define the character with subleading
behavior as $q \rightarrow 0$ to have leading behavior $q^{-c_{\rm eff}/24+ h_{\rm eff}}$ then the values of $h_{\rm eff}$ are
\begin{align}
h_{\rm eff}(X)=\{1/5,1/4,1/3,2/5,1/2,3/5,2/3,3/4,4/5 \}
\end{align}
and these numbers are exactly the entries not equal to $0$ or $1$ in the Farey
sequence of order $5$ \footnote{Perhaps $\frac{0}{1}$ and $\frac{1}{1}$ in this sequence should be identified with the trivial CFT
and affine $E_8$ at level $1$.}.
\be
F_5 = \left\{ \frac{0}{1}, \frac{1}{5}, \frac{1}{4}, \frac{1}{3}, \frac{2}{5}, \frac{1}{2}, \frac{3}{5}, \frac{2}{3}, \frac{3}{4}, \frac{4}{5}, \frac{1}{1} \right\}
\ee
We do not know an explanation for this observation. 

For $n=2$ and $\ell=2$ the modular Wronskian must have ${\rm ord}_\omega(W)= 1$ which implies it is proportional to $E_4$. We then have
$\phi_0 \sim E_6/E_4$ and $\phi_1 \sim E_4$. The resulting equation appears to have two independent parameters but in fact this can be reduced
to a single parameter using the indicial equation as in \cite{Hampapura:2015cea}. The solutions that might correspond to RCFT characters were
analyzed in \cite{naculich, Hampapura:2015cea}, see for example Table 1 in \cite{Hampapura:2015cea}. We will show in section \ref{subsec:mlderel}
that all of these characters can be simply obtained as Hecke images of the $\ell=0$ characters described above. In \cite{Hampapura:2015cea} no solutions
were found with $\ell=3,4,5$. We will comment later on the significance of this result. 

We now turn to third order MLDE. 
For $n=3$ and $\ell=0$ the MLDE involves two free parameters $\mu_1$ and $\mu_2$, and  takes the form
\begin{align} \label{thirdmlde}
\left(D^3 -\frac{1}{2}E_2D^2+\Big(\frac{1}{2}(DE_2)+\big(\frac{1}{18} - \frac{\mu_1}{4}\big)E_4D- \frac{\mu_2}{8}E_6 \Big) \right) f=0.
\end{align}
It can be solved by the same method as in the second order case.
There will be three solutions of Frobenius form:
\begin{align} \label{frobform}
\chi_i=q^{\alpha_i}\sum_{n=0}^\infty c_i(n)\, q^n, 
\qquad i=0,1,2,
\end{align}
with  $c_0(0)=1$ for the vacuum character. The indicial equation for the coefficients in the $q$ expansion gives
\begin{align}
\mu_1=\frac{2}{9}-4(\alpha_0\alpha_1+\alpha_0\alpha_2+\alpha_1\alpha_2),
\qquad
\mu_2=8\alpha_0\alpha_1\alpha_2.
\end{align}
There is to our knowledge no complete list of solutions of this third order equation of RCFT character type, that is with
positive integer coefficients $c_i(n)$ in the $q$ expansion (\ref{frobform}).  A number of solutions with $\ell=0$ were studied and conjectures
made in \cite{kaneko}. In \cite{Hampapura:2016mmz} solutions with $\ell=0$ that could correspond to RCFTs with no dimension one operators
and hence no Kac-Moody symmetry were classified. We will see that many of these solutions can be obtained and some of their properties understood by viewing
them as Hecke images of solutions corresponding to known RCFTs. We are not aware of a study of three character theories with $\ell>0$ but we will show
that many such characters can be constructed using Hecke operators.

In order to give examples of Hecke relations, in section \ref{sec:Applications} we will relate a number of RCFT with three independent
characters to the characters of three simpler RCFT which we use to illustrate the Hecke relations. These simpler models are
the Ising model, that is the minimal model $\mathsf{M}(3,4)$, the minimal model $\mathsf{M}(2,7)$ and the tensor product of the
Yang-Lee model of minimal model $\mathsf{M}(2,5)$ with itself. 
In the following we give the characters of these three models with a subscript  added to each character indicating the (effective) weight of the
corresponding primary field.

The characters of the Ising model are given by
\begin{align}
\begin{split}
\chi^I_0 &=\frac{1}{2}\left( \sqrt{\frac{\theta_3 (\tau)}{\eta (\tau)}} +\sqrt{\frac{\theta_4 (\tau)}{\eta (\tau)}} \right)
=q^{-1/48}(1 + q^2 + q^3 + 2 q^4 + 2 q^5 +\cdots) ,
\\
\chi^I_{1/2} &=\frac{1}{2}\left( \sqrt{\frac{\theta_3 (\tau)}{\eta (\tau)}} -\sqrt{\frac{\theta_4 (\tau)}{\eta (\tau)}} \right)
=q^{23/48}(1 + q + q^2 + q^3 + 2 q^4 + 2 q^5+\cdots), 
\\
\chi^I_{1/16} &=\frac{1}{\sqrt{2}}\sqrt{\frac{\theta_2 (\tau)}{\eta (\tau)}} 
=q^{1/24}(1 + q + q^2 + 2 q^3 + 2 q^4 + 3 q^5+\cdots) \, .
\end{split}
\end{align}

In the minimal model $M(2,7)$, the three characters are \cite{Hikami:Spherical_Seifert}
\begin{align}
\begin{split}
\chi_{0}(\tau)&=\frac{q^{1/56}}{\eta(\tau)}
\prod_{n=1}^{\infty} (1-q^{7n})(1-q^{7n-4})(1-q^{7n-3}) \\
&= q^{-1/42}(1+q+ 2q^2+ 2q^3+ 3q^4+\cdots) , \\
\chi_{1/7}(\tau)&=\frac{q^{9/56}}{\eta(\tau)}
\prod_{n=1}^{\infty} (1-q^{7n})(1-q^{7n-5})(1-q^{7n-2}) \\
&= q^{5/42}(1+q+ q^2+ 2q^3+ 3q^4 + \cdots)  ,\\
\chi_{3/7}(\tau)&=\frac{q^{25/56}}{\eta(\tau)}
\prod_{n=1}^{\infty} (1-q^{7n})(1-q^{7n-6})(1-q^{7n-1}) \\
&= q^{17/42}(1+q^2+q^3+ 2q^4 + \cdots) \, .
\end{split}
\end{align}

Finally, the characters of the tensor product of the Yang-Lee  model with itself are readily obtained as products of Yang-Lee characters (note the representation 
with $h=1/5$ appears with multiplicity two in the tensor product theory, so the tensor product theory has four characters, but only three independent characters). 
\begin{align}
\begin{split}
\chi_0 =&   \chi^{YL}_0 (\tau)^2, \\
\chi_{1/5} =&  \chi^{YL}_0 (\tau)\chi^{YL}_{1/5}(\tau) , \\
\chi_{2/5} =&   \chi^{YL}_{1/5} (\tau)^2 .
\end{split}
\end{align}

In \cite{Hampapura:2016mmz}  the following solutions are presented, which although not all  linked to specific RCFTs,  are solutions with positive integer coefficients and
thus possible characters of RCFTs or intermediate vertex subalgebras.  They are distinguished by the fact that the characters correspond to a theory with no dimension one operators,
that is no Kac-Moody symmetry. The first solution consists of the characters of the Baby Monster VOA constructed in \cite{hoehn}
with $q$ expansions
\begin{align}
\begin{split}
\chi_{VB^\natural_{(0)}} &= q^{-47/48} \left( 1 + 96256 q^2 + 9646891 q^3  + \cdots \right), \\
\chi_{VB^\natural_{(1)}} &= q^{25/48} \left( 4371 +1143745 q + 64680601 q^2 + \cdots \right) ,\\
\chi_{VB^\natural_{(2)}} &= q^{23/24} \left( 96256 + 10602496 q + 420831232 q^2 + \cdots \right).
\end{split}
\end{align}
These are characters of a RCFT with central charge $c=47/2$ and conformal weights $h_1=3/2$ and $h_2=31/16$ and arise as solutions of eqn.(\ref{thirdmlde}) with
$\mu_1=2315/176 $ and $\mu_2=-27025/6912 $. The second solution would correspond to
a RCFT with central charge $c=236/7= 59*4/7$ and conformal weights $h_1=16/7$ and $h_2=17/7$  and is a solution of eqn.(\ref{thirdmlde}) with
$\mu_1=461/63$ and $\mu_2=-93869/9261$. The characters are given by
\begin{align}
\begin{split}
\chi^{236/7}_0 &= q^{-59/42} \left(1 + 63366 q^2 + 46421200 q^3 + 5765081101 q^4 + \cdots \right)  \, , \\
\chi^{236/7}_{16/7} &= q^{37/42} \left( 715139+257698784 q+24078730130 q^2 + 1120165372784 q^3 + \cdots     \right) \, ,\\
\chi^{236/7}_{17/7} &= q^{43/42} \left(848656+  232637826 q + 19201964416 q^2 + 828747166732 q^3        + \cdots \right) \, .
\end{split}
\end{align} 
The third solution would correspond to a RCFT with central charge $c=164/5=41*4/5$ and conformal weights $h_1=11/5$ and $h_2=12/5$ and is a solution to eqn.(\ref{thirdmlde})
with $\mu_1=1571/225$ and $\mu_2=-1271/135$.
The characters are given by
\begin{align}
\begin{split}
\chi^{164/5}_0 &=  q^{-41/30} \left(1+90118 q^2+53459408 q^3 + 5940658961 q^4 + \cdots \right)  ,  \\
\chi^{164/5}_{11/5} &= q^{5/6} \left( 254200+92624125 q+ 8457234000 q^2 + 381896416750 q^3 + \cdots \right), \\
\chi^{164/5}_{12/5} &= q^{31/30} \left(615164 + 152560672 q + 11717226984 q^2 + 476717427120 q^3 + \cdots \right) .
\end{split}
\end{align}

We will show in section \ref{sec:Applications} that the characters of these three examples are images under Hecke operators
$\mathsf{T}_{47}, \mathsf{T}_{59}$ and $\mathsf{T}_{41}$ of the characters of minimal models $\mathsf{M}(3,4)$, $\mathsf{M}(2,7)$ and $\mathsf{M}(2,5)^{\otimes 2}$ respectively.

\section{Scalar Hecke Operators}\label{sec:ScalarHecke}

In this section we give a brief review of standard Hecke operators for $\Gamma=SL(2,\IZ)$ and the principal congruence subgroup
$\Gamma(N)$.  For $SL(2,\IZ)$ this is standard material and may be familiar from discussions in \cite{serre, zagier}. Hecke operators for congruence subgroups are probably less familiar to
physicists. References we found useful include \cite{diashur, Rankin, shimura}. 

Modular forms for $SL(2,\IZ)$ of weight $k$ can
be viewed either as holomorphic functions of $\tau \in \IH$ which transform as
\be
f \left( \frac{a \tau+b}{c \tau+d} \right) = (c \tau+d)^k f(\tau)
\ee
under modular transformations, or as functions of rank two lattices $L$ which obey
\be
F( \lambda L) = \lambda ^{-k} F(L)
\ee
under rescaling of the lattice $L$.  The two are related via $F(\IZ \omega_1 + \IZ \omega_2) = \omega_2^{-k} f(\omega_1/\omega_2)$.

\subsection{Hecke Operators for \texorpdfstring{$SL(2,\IZ)$}{SL(2,\IZ)}}

The lattice viewpoint gives a simple definition of the Hecke operator $T_n$ acting on modular forms of weight $k$ for $SL(2,\IZ)$.  For $n \in \IZ^+$, $T_n$ simply sums a function $F(L)$ over sublattices $L'$ of index
$n$ in $L$:
\be
(T_n F)(L) = \sum_{\substack{L' \subset L \\ |L/L'|=n}} F(L')
\ee
If $\omega_1, \omega_2$ are a basis of $L$ then we can generate any sublattice $L'$ of index $n$ as the lattice with basis vectors
\begin{align}
\begin{split}
\omega'_1 &= a \omega_1 + b \omega_2 \\
\omega'_2 &= c \omega_1 + d \omega_2
\end{split}
\end{align}
with $a,b,c,d \in \IZ$ and $ad-bc=n$. Write $\CM_n(\IZ)$ for the set of two by two matrices with entries in $\IZ$ and determinant $n$.

We can now translate the above definition into Hecke operators acting on weight $k$ modular forms as functions of $\tau$. We first define
an action of $ \mu \in \CM_n(\IZ)$ on such forms by
\be\label{def:f|}
(f|_k \mu)(\tau) = \frac{({\rm det}(\mu))^{k/2}}{(c \tau+d)^k} f(\mu \tau)
\ee
with 
\be
\mu= \begin{pmatrix} a & b \\ c & d \end{pmatrix} \, , \qquad \mu \tau= \frac{a \tau +b}{c \tau+d} \, .
\ee
The modular group $\Gamma$ acts on $\CM_n(\IZ)$ by left multiplication and a set of representatives for $\Gamma \backslash \CM_n(\IZ)$ is given
by
\be
\Delta^{(n)} =  \left \{ \begin{pmatrix} a & b \\ 0 & d \end{pmatrix} \Big| ad=n, d>0, 0 \le b < d \right \} \, .
\ee

The action of the $n$th Hecke operator is then
\be\begin{split} \label{eqn:scalarhecke}
(T_n f)(\tau) &=  n^{k/2-1} \sum_{ \delta \in \Delta^{(n)}} (f|_k \delta)(\tau)\\
&= n^{k-1} \sum_{\substack{ad=n \\ a,d>0}} \frac{1}{d^k} 
\sum_{b~~(\text{mod }d)} f\Big(\frac{a\tau+b}{d}\Big).
\end{split}\ee 
Hecke operators $T_m$, $T_n$ for $m,n$ relatively prime obey
\be \label{Tm_Tn}
T_m T_n f = T_{mn} f ,
\ee
and  for $n=p^{m+1}$ with $p$ prime and $m \ge 1$
\be \label{T_p^n}
T_{p^{m+1}} f= T_p T_{p^m} f - p^{k-1} T_{p^{m-1}} f \, .
\ee
so it suffices to know the action of $T_p$ on weight $k$ forms $f$ for $p$ prime. For prime $p$ we have
\be
\Delta^{(p)} =  \left \{ \begin{pmatrix} p & 0 \\ 0 & 1 \end{pmatrix}, \begin{pmatrix} 1 & b \\ 0 & p \end{pmatrix}  \Big| 0 \le b < p \right \} \, .
\ee
To prove that $T_p f$ is again a modular form we note that the transformations $T \tau$ and $S \tau$ acting on $(T_p f)(\tau)$ are equivalent to replacing
the sum over elements $\delta$ of $\Delta^{(p)}$ with a sum over $\delta T$ or $\delta S$ respectively. Then we note that for each $\delta \in \Delta^{(p)}$, $\delta T$ and $ \delta S$
are equivalent to $\gamma \delta'$ for some $\gamma \in \Gamma$ and $\delta' \in \Delta^{(p)}$ and that each $\delta' \in \Delta^{(p)}$ arises this way. Thus acting
on $\Gamma \backslash \CM_p(\IZ)$ the right action of $T$, $S$ simply permute the elements of the representatives in $\Delta^{(p)}$ and hence leave the sum defining
$T_p f$ invariant.

The above is easily translated into an action on the coefficients of the $q$ expansion of a modular form. If $f= \sum_n a(n) q^n$ is a modular form of weight $k$, then $(T_p f)(\tau)=\sum_n a^{(p)}(n)q^n$ with the Fourier coefficients
\begin{align}
a^{(p)} (n)=
\begin{cases} p^k a(pn)
& \text{if  } p \nmid n , \\
p^{k-1} \big(p\,a(pn) + a(n/p) \big)
&  \text{if  } p | n .
\end{cases} 
\end{align}
Note that if $a(n) \in \IZ$ then $a^{(p)}(n) \in \IZ$ for $k \ge 1$. Since most mathematical applications of Hecke operators use the action on weight $k$ modular forms for $k \ge 1$, this normalization
is convenient. However we wish to define Hecke operators on vector-valued weight zero modular functions and for this case this standard normalization does not
preserve integrality of the coefficients. As a result the definition of a new class of Hecke operators we give later replaces the factor of $n^{k-1}$ in eqn.(\ref{eqn:scalarhecke}) by $1$
when acting on weight zero modular functions.

\subsection{Hecke Operators for  \texorpdfstring{$\Gamma(N)$}{Gamma(N)}}

To extend the treatment of Hecke operators for $SL(2,\IZ)$ to Hecke operators for the principal congruence subgroup $\Gamma(N)$ it is useful to
recast the treatment of Hecke operators in terms of a double coset construction. The general setup, as discussed say in \cite{diashur} involves two
congruence subgroups $\Gamma_1, \Gamma_2$ of $\Gamma=SL(2,\IZ)$ and an element $\alpha \in GL^+(2, \IQ)$. The set
\be
\Gamma_1 \alpha \Gamma_2 = \{ \gamma_1 \alpha \gamma_2 | \gamma_1 \in \Gamma_1, \gamma_2 \in \Gamma_2 \}
\ee
is a double coset in $GL^+(2, \IQ)$ and can be used to define a map from weight $k$ modular forms for $\Gamma_1$ to weight $k$ modular forms
for $\Gamma_2$. This map acting on a weight $k$ modular form $f$ for $\Gamma_1$ is
\be
f[\Gamma_1 \alpha \Gamma_2]_k = \sum_j f |_k \delta_j \, .
\ee
Here we have extended the slash operator from $\CM_n(\IZ)$ to $GL^+(2,\IQ)$ in the obvious way and the double coset $\Gamma_1 \alpha \Gamma_2$ is
written as a disjoint union $\bigcup_j \Gamma_1 \delta_j$ over orbit representatives $\delta_j$. 

The double coset machinery can be used to define Hecke operators $T_p$ for modular forms for the principal congruence subgroup $\Gamma(N)$. Through out this subsection, we restrict to $p$ prime. The extension to Hecke operators $T_n$ for $n$ relatively prime to $N$ but not necessarily
prime is presented in Appendix \ref{app:Hecke}.

Following the discussion and notation of \cite{Rankin} we take $\Gamma_1=\Gamma_2= \Gamma(N)$ and
\be
\alpha= J_p = \begin{pmatrix} 1 & 0 \\ 0 & p \end{pmatrix} 
\ee
and consider the double coset $(J_p)=\Gamma(N) J_p \Gamma(N)$. We also introduce
\be
[\![J_p]\!]= \left\{ \begin{pmatrix} a & b \\ c & d \end{pmatrix} \big| a,b,c,d \in \IZ, ad-bc=p, \begin{pmatrix} a & b \\ c & d \end{pmatrix} \equiv J_p ~{\rm mod}~N  \right\}
\ee
which is a union of double cosets. In \cite{Rankin} it is shown that
\be \label{rankin1}
[\![J_p]\!]=\Gamma(N) \cdot {\cal T}_p^+ = {}^+ {\cal T}_p \cdot \Gamma(N)
\ee
and that
\be \label{rankin2}
{\cal M}_p(\IZ) = \Gamma(1) \cdot {\cal T}_p^+ = {\cal T}_p^+ \cdot \Gamma(1)
\ee
where $\Gamma(1)=SL(2, \IZ)$, ${}^+ {\cal T}_p$ is a left transversal whose form will not be needed and ${\cal T}_p^+$ is a right transversal given by
\be\label{rep_gen}
{\cal T}_p^+=\left\{
\sigma_p \beta_p;   U_b, 0\leq b \leq p-1 \right\},
\ee
where
\be
\beta_p = \begin{pmatrix} p & 0 \\ 0 & 1 \end{pmatrix}\, ,  \qquad U_\nu  = \begin{pmatrix}1 & \nu N \\ 0 & p \end{pmatrix} \, ,
\ee
we write $\bar p$ for the multiplicative inverse of $p$ in $(\IZ/N \IZ)^\times$
and we denote by $\sigma_p$ the preimage of
\be
\begin{pmatrix} \bar p & 0 \\ 0 & p \end{pmatrix}
\ee
under the mod $N$ map from $SL(2, \IZ)$ to $SL(2, \IZ/N \IZ)$. The set of matrices ${\cal T}_p^+$ can be regarded as a generalization to $\Gamma(N)$  of the $\Delta^{(p)}$ that appeared in the description of Hecke operators for $SL(2,\IZ)$ and in the following we will also denote this set of matrices by $\Delta^{(p)}_N$ in order to emphasize
the dependence on $N$.

The Hecke operator $T_p$ acting on a weight $k$ modular form for $\Gamma(N)$ is then defined to be
\be \label{heckeNdef}
(T_p f)(\tau)= \sum_{ \delta \in \Delta_N^{(p)}} (f|_k \delta)(\tau) \, .
\ee
To show that $(T_p f)$ is again a modular form of weight $k$ for $\Gamma(N)$ we note that taking $\tau \rightarrow \gamma \tau$ with
$\gamma \in \Gamma(N)$ replaces the sum over $\delta$ in \eqref{heckeNdef} by a sum over $\delta \gamma$ but since $[\![J_p]\!] \gamma = [\![J_p]\!]$ from
eqn. (\ref{rankin1}), the sum over coset representatives and hence $T_p f$ is unchanged, that is $T_p f$ is invariant under the weight $k$ slash operator
for $\Gamma(N)$ if $f$ is.

We note that $\Gamma(N)$ does not contain $T$, but does
contain $T^N$ and that therefore modular forms for $\Gamma(N)$ are invariant under $\tau \rightarrow \tau+N$ which implies that the $q$ expansion of a modular form for $\Gamma(N)$ can be written in the form
\be
f(\tau) = \sum_n b(n) q^{n/N} \, .
\ee
However, because $(T_p f)(\tau) $ includes the term
\be
(f|_k \sigma_p \beta_p) (\tau) = p^{k-1}(f |_k \sigma_p)(p \tau)
\ee
there is no simple way to write the action of Hecke operators on modular forms for $\Gamma(N)$ in terms of an action on their Fourier coefficients $b(n)$. This is dealt with in the mathematical literature by introducing modular forms transforming according to a character $\chi$ such that there is
a good action of the Hecke operators on their Fourier coefficients. 
We will see in the following section that this can also be remedied when we have level $N$ vector-valued modular functions $f$  with respect to a representation $\rho$ by defining a variant of $T_p$ that leads to mixing of the vector components of $f$ and does have a simple action on the Fourier coefficients of the components $f_i$ of $f$.  The relation between the two formulations is that the representation by which the components transform
can be diagonalized over $\IC$ to give modular forms transforming according to specific combinations of Dirichlet characters.

\section{Hecke Operators for RCFT Characters}\label{sec:HeckeRCFT}

It is not obvious how one should extend the definition of Hecke operators to vector-valued modular forms. The basic problem is that the definition of Hecke operators involves a sum over matrices
in $\CM_n(\IZ)$ while vector-valued modular forms transform under modular transformations according to a representation $\rho$ of $SL(2,\IZ)$ and it is not clear how to extend
this representation to a large enough group to include the elements of $\CM_n(\IZ)$ required to define Hecke operators. In the mathematics literature some ways to deal with this have been discussed in \cite{brunier, mwr}. The paper \cite{brunier} constructs Hecke operators for vector-valued forms transforming according to Weil representations of either the modular group
(for integer weight modular forms) or the metaplectic group (for half integer weight forms). They do this by extending a representation of $SL(2, \IZ/N\IZ)$ to a representation
of a group $Q(N)$  isomorphic to $SL(2, \IZ/N \IZ) \times (\IZ/N \IZ)^\times$ and defining Hecke operators for elements $(M,r) \in SL(2, \IZ/N \IZ) \times (\IZ/N \IZ)^\times$ with
$\det M= r^2~{\rm mod}~N$. The Hecke operators defined
in \cite{mwr} take a representation $\rho$ of $SL(2,\IZ)$ to a representation on a complex vector space with basis elements labelled by the set of orbit representatives of a double coset. 
Our Hecke operators in contrast take representations of dimension $n$ of
$SL(2, \IZ)$ to representations of the same dimension and are defined without much of the mathematical machinery employed in \cite{brunier, mwr}. Nonetheless, since
the representations of $SL(2,\IZ)$ that appear in RCFT can be recast in the language of Weil representations (see e.g. \cite{eholzerI, eholzerII}) it is quite likely that
the Hecke operators we define can be recast in the language of \cite{brunier, mwr}. Hecke operators on vector-valued modular forms also appear in \cite{diacon}.

\subsection{Definition and Properties}

In spite of the above difficulties, it is rather trivial to define Hecke operators acting on level $N$ modular functions with respect to a representation $\rho$, taking
a modular function relative to the representation $\rho$ to one relative to a representation $\rho^{(p)}$, also with $\Gamma(N) \subset {\rm ker}(\rho^{(p)})$.  We simply note that the first term  in the set of representatives of eqn.(\ref{rep_gen}) acting on the components $f_i$  gives $f_i(\sigma_p  p \tau) =\sum_j \rho(\sigma_p)_{i j} f_j(p \tau)$ which mixes the components of $f$ and  does not require extending the definition of 
$\rho$ to a larger group.  To be slightly more precise we now use the action of the orbit representatives in eqn.(\ref{rep_gen})  on  weight zero vector-valued modular functions relative to a representation $\rho$ whose
kernel contains $\Gamma(N)$ to define the Hecke operator for $p$ prime
\be \label{scalarHeckeN}
(\mathsf{T}_p f)_i(\tau) =  \sum_{ \delta \in \Delta_N^{(p)}} f_i(\delta \tau) = \sum_j \rho_{ij}(\sigma_p) f_j(p \tau) +\sum_{b=0}^{p-1} f_i\left(\frac{\tau+bN}{p}\right) .
\ee
since by the definition of a vector-valued weight zero modular function $f$ relative to the representation $\rho$ we have
\be
f_i( \gamma \tau)=  \sum_j \rho(\gamma)_{ij} f_j(\tau) 
\ee
for $\gamma \in SL(2, \IZ)$. 
The normalization of $\mathsf{T}_p$ differs from $T_p$ for scalar modular forms encountered before in order to preserve the integrality of coefficients.
One essential feature of these Hecke operators is that they do lead to a nice action on the Fourier coefficients.  If
\be
f_i(\tau) = \sum_n b_i(n) q^{n/N}
\ee
then using
\be
\sum_{b=0}^{p-1} e^{2 \pi i bn/p} = \begin{cases} p & p|n \\ 0 & \text{otherwise} \end{cases}
\ee
in eqn.(\ref{scalarHeckeN}) gives
\be
(\mathsf{T}_p f)_i(\tau) = \sum_n b^{(p)}_i(n) q^{n/N}
\ee
with
\be \label{HeckeRcftCoef}
b^{(p)}_i(n) = \begin{cases}  p b_i(pn) &  p \nmid n \\ p b_i(pn) + \sum_j \rho_{ij}(\sigma_p) b_j(n/p) & p|n \end{cases}
\ee

We now prove that  if $ f$ is a level $N$ vector-valued modular function with representation $\rho$ then $(\mathsf{T}_p f)$ is a level $N$ modular function with representation $\rho^{(p)}$ of the form claimed earlier.  First consider the action of $\gamma \in \Gamma(N)$. The fact that each component of $(\mathsf{T}_p f)$ is a modular form for $\Gamma(N)$ follows from our discussion of Hecke operators for $\Gamma(N)$ and the fact
that $\Gamma(N) \subset {\rm ker}(\rho)$. To determine the representation $\rho^{(p)}$ it suffices to determine it on the generators $T, S$ of $SL(2,\IZ)$. Because of eqn.(\ref{rankin2}),
for each $\gamma \in \Gamma $ and $\delta \in \Delta_N^{(p)}$ we have $\delta \gamma = \gamma' \delta' $ for some $\gamma' \in \Gamma$ and $\delta' \in \Delta_N^{(p)}$. We want to
determine $\gamma'$ for $\gamma=T \text{ and } S$ modulo the action of $\Gamma(N)$.

In particular we show that
\begin{align}
\begin{split}
\Delta_N^{(p)} \circ T &= T^{\bar p} \circ \Delta_N^{(p)} ,\\
\Delta_N^{(p)} \circ S &= \sigma_p S \circ \Delta_N^{(p)} \, .
\end{split}
\end{align}
which shows that the Hecke image $\mathsf{T}_p f$ transforms under $SL(2,\IZ)$ according to the representation $\rho^{(p)}$ with
$\rho^{(p)}(S) = \rho(\sigma_p S)$ and $\rho^{(p)}(T)= \rho(T^{\bar p})$.

We first consider the modular transformation $T:\tau \rightarrow \tau+1$ and its right action
on the second set of representatives in eqn. \eqref{rep_gen}
\begin{align}
U_{\nu} =
\begin{pmatrix}
1 & \nu N \\0 & p
\end{pmatrix},
\end{align}
where $\nu$ runs over a complete set of residues mod $p$.
We have 
\begin{align}
U_{\nu} T = T^{\bar p}U_{\nu'(\nu)} ~,
\end{align}
with 
\begin{align}
\nu'(\nu) =\nu- \frac{p\bar p -1}{N}.
\end{align}
Since $p \bar p =1 ~{\rm mod}~N$, when $\nu$ runs over a complete set of residues ${\rm mod}~p$ so does $\nu'$.

To deal with the first representative in eqn. \eqref{rep_gen} 
we note that
\be
\sigma_p \beta_p T = \gamma' \sigma_p \beta_p
\ee
with
\be
\gamma' = \sigma_p \begin{pmatrix} 1 & p \\ 0 & 1 \end{pmatrix} \sigma_p^{-1}
\ee
and that $\gamma' \in SL(2, \IZ)$ since $\sigma_p \in SL(2, \IZ)$.  Furthermore working mod $N$ we have
\be
\gamma'= \begin{pmatrix} \bar p & 0 \\ 0 & p \end{pmatrix} \begin{pmatrix} 1 & p \\ 0 & 1 \end{pmatrix} \begin{pmatrix} p & 0 \\ 0 & \bar p \end{pmatrix} = \begin{pmatrix} 1 & \bar p \\ 0 & 1 \end{pmatrix} \ee
so $\gamma'= T^{\bar p}$ up to the action of $\Gamma(N)$. 
This establishes that
\begin{align}
\Delta_N^{(p)} \circ T= T^{\bar p} \circ \Delta_N^{(p)} ,
\end{align}
and verifies the modular representation of $T$ for the Hecke image $\mathsf{T}_p f$,
\begin{align}\label{rho^p_T}
\rho^{(p)}(T)  = \rho(T^{\bar p}).
\end{align}

Next we consider the transformation $S:\tau\rightarrow -1/\tau$ and first deal with the action on the representatives $U_\nu$ for $\nu \neq 0 ~{\rm mod} ~p$.
For each $\nu\neq 0 \text{ mod }p$, we have
\be
U_\nu S = \gamma'_\nu U_{\nu'}
\ee
with
\begin{align}
\gamma'_{\nu}\equiv                                                                                                                                                                                       
U_{\nu} S U^{-1}_{\nu'}=
\begin{pmatrix}
N\nu & -(1+N^2\nu\nu')/p \\
p & -N\nu'
\end{pmatrix}.
\end{align}
To find a $\nu'$ such that $\gamma'_\nu \in SL(2, \IZ)$ we note we must have
\begin{align}
1+N^2\nu\nu'=0\quad  \text{ mod }p
\end{align}
which has the solution $\nu'= -(N^2\nu)^{p-2}$ by Fermat's little theorem. With this choice of $\nu'$, $\gamma'_\nu$ has integral entries and determinant
one and so is in $SL(2, \IZ)$. One then checks that $\gamma'_\nu S^{-1}$ is congruent to $\sigma_p$ mod $N$ so we can take $\gamma'_\nu = \sigma_p S$.
Since $p$ is a prime number, we have that $\{\nu'(\nu)| 1\leq \nu \leq p-1\}$ is a permutation of the elements of $(\IZ/p\IZ)^{\times}$.

There are two representatives in $\Delta_N^{(p)} $ left to deal with, i.e.
\begin{align}
U_0 \quad \text{and} \quad 
\sigma_p
\begin{pmatrix}
p & 0 \\ 0 & 1
\end{pmatrix}.
\end{align}
One easily checks that
\begin{align}
\begin{split}
U_0  S &=\gamma'_0 \sigma_p
\begin{pmatrix}
p & 0 \\ 0 & 1
\end{pmatrix} ,\\
\sigma_p
\begin{pmatrix}
p & 0 \\ 0 & 1
\end{pmatrix}S
&=\gamma'_{\sigma}U_0,
\end{split}
\end{align}
with $\gamma'_0$ and $\gamma'_{\sigma}$ in $SL(2, \IZ)$ and equal to $\sigma_p S$ up to the action of $\Gamma(N)$. 

Having verified
\begin{align}
\Delta_N^{(p)} \circ S= \sigma_p S \circ \Delta_N^{(p)} ,
\end{align}
we find the modular representation of $S$ for the Hecke image $\mathsf{T}_p f$,
\begin{align}\label{rho^p_S}
\rho^{(p)}(S)  = \rho(\sigma_p S).
\end{align}
and thus verify eqn. (\ref{newrep}). 

For simplicity we have restricted our discussion to $p$ prime, but Hecke operators $\mathsf{T}_n$ for $n$ relatively prime to $N$ can
also be constructed, see Appendix \ref{app:Hecke}. 

\subsection{Relation to Galois Symmetry}
\label{subsec:Relation to Galois symmetry}

In section \ref{subsec:Galois} we discussed the Frobenius elements $f_{N, \ell}$ for $\ell \in (\IZ/ N \IZ)^\times$ which act on the representation matrices $\rho(T), \rho(S)$ as
\be
f_{N, \ell} \big( \rho(T) \big) = \rho(T)^\ell, \qquad
f_{N, \ell}\big(\rho(S)\big)= G_\ell \,\rho(S) \, .
\ee
On the other hand, we have seen that the Hecke operators $\mathsf{T}_p$ for $p$ relatively prime to $N$ also act to produce a new representation $\rho^{(p)}$ from $\rho$. Here we establish that these two actions are related by
\be \label{Heckefrob}
f_{N, p}\big( \rho(S) \big) = \rho^{(\bar p)}(S), \qquad 
f_{N, p}\big(\rho(T) \big)= \rho^{(\bar p)}(T) \, .
\ee
Thus the Hecke operators $\mathsf{T}_{\bar p}$ agree with the Galois action $f_{N, p}$ on the modular representation $\rho$ but extend it to an action on the characters
of the RCFT rather than just an action on the modular representation.

The main result we require to show this is Theorem 1 of \cite{Bantay:2001ni}:
\begin{align}\label{T_l^2}
 G_{\ell} \,\rho(T)\, G_{\ell}^{-1}=\rho (T)^{\ell^2} , \qquad
\ell \in (\mathbb{Z}/N\mathbb{Z})^{\times} \, .
\end{align}

Writing the above as $\rho(T) = G_\ell^{-1} \rho(T)^{\ell^2} G_\ell$, taking the $\bar \ell$ power of both sides and using the fact that
$\rho(T)^N=1$ leads to the equivalent relation
\begin{align}\label{Tp Tpbar}
G_p^{-1}\, \rho(T)^p \, G_p= \rho(T)^{\bar p}, \qquad
p \in (\mathbb{Z}/N\mathbb{Z})^{\times} \, .
\end{align}

To establish eqn.(\ref{Heckefrob}) we first note that $C= \rho(S)^2$ is invariant under the Frobenius map $f_{N,p}$ since 
\begin{align}
\rho(S) \rho(S) \rightarrow \rho(S) \, G_p^{-1} G_p \,\rho(S)= \rho(S)^2.
\end{align}
We now apply the Frobenius map $f_{N, p}$ to both sides of the equation:
\begin{align}\label{C=ST^3}
\big( \rho(S) \rho(T) \big)^3 = C ,
\end{align}
to find
\begin{align}
\rho(S)\, G_p^{-1}\, \rho(T)^p \cdot
G_p\,\rho (S)\, \rho(T)^p \cdot
\rho(S)\, G_p^{-1}\, \rho(T)^p
=C=\rho(S)^2 ,
\end{align}
which using the identity eqn.\eqref{Tp Tpbar},
becomes
\begin{align}
G_p =  \rho(T)^p\, \rho (S)^{-1}\, \rho(T)^{\bar p} \, \rho (S) \, \rho(T)^p \, \rho (S) 
=\rho \big( T^p\, S^{-1}\, T^{\bar p} \, S \, T^p \, S \big) \, .
\end{align}
An explicit computation of the argument gives
\begin{align}
 T^p\, S^{-1}\, T^{\bar p} \, S \, T^p \, S  =
\left(
\begin{array}{cc}
 \left(1-p \bar{p}\right) p+p & p \bar{p}-1 \\
 1-p \bar{p} & \bar{p} \\
\end{array}
\right)
\equiv 
\begin{pmatrix}
p & 0 \\ 0 & \bar p
\end{pmatrix}
\text{  mod }N ,
\end{align}
which establishes that $ T^p\, S^{-1}\, T^{\bar p} \, S \, T^p \, S \in SL(2, \IZ)$ is the preimage of $\sigma_{\bar p}$ under the mod $N$ map from $SL(2,\IZ)$ to $SL(2,\IZ/N \IZ)$
and thus that $G_p = \rho( \sigma_{\bar p})$. This establishes the first relation in eqn.(\ref{Heckefrob}) and the second relation follows trivially from the definitions since $\rho(T)$ is diagonal. 
%\end{proof}

It will be useful later to state a few results concerning the possible values of the matrix $\rho(\sigma_p)$ that appears in the Hecke transformation of coefficients, eqn.(\ref{HeckeRcftCoef}),
and in the relation between $\rho^{(p)}(S)$ and $\rho(S)$. We note that $\rho(\sigma_p)$ is periodic in $p$ with period $N$, the conductor of the RCFT. In addition we have
$\rho(\sigma_{p_1}) \rho(\sigma_{p_1}) = \rho(\sigma_{p_1 p_2})$. Thus $\rho(\sigma_p)$ provides a representation of the multiplicative group of units, $(\IZ/N \IZ)^\times$. Since this group
is Abelian, all representations are one-dimensional over $\mathbb{C}$. Therefore, working over $\mathbb{C}$ we can always decompose the representation provided by $\rho(\sigma_p)$
into a finite sum of Dirichlet characters mod $N$. As an example consider Hecke images $\mathsf{T}_p \chi^{YL}$ of the Yang-Lee model characters with $p \in (\IZ/60 \IZ)^\times$. One easily finds that
\begin{align} \label{ylsigp}
\begin{split}
\rho^{YL}(\sigma_p) &= \begin{pmatrix} 1 & 0 \\ 0 & 1 \end{pmatrix} ~{\rm for}~p=1,19,41,59 ~{\rm mod~}60, \\
                          &=  \begin{pmatrix} -1 & 0 \\ 0 & -1 \end{pmatrix} ~{\rm for}~p=11,29,31,49 ~{\rm mod~}60, \\
                          &=  \begin{pmatrix} 0 & -1 \\ 1 & 0 \end{pmatrix} ~{\rm for}~p=7, 13, 47, 53  ~{\rm mod~}60,  \\
                          &=   \begin{pmatrix} 0 & 1 \\ -1 & 0 \end{pmatrix} ~{\rm for}~p=17, 23, 37, 43  ~{\rm mod~}60\, .
\end{split}                          
\end{align}
and finds that $\rho^{YL}(\sigma_p)$ can be diagonalized over $\mathbb{C}$ into the direct sum $X_{10}(p)+X_{12}(p)$ where $X_{10}$ and $X_{12}$ are Dirichlet characters mod $60$ which may be defined through their values on the generators $7,19,11$ of $(\IZ/60 \IZ)^\times \simeq C_4 \times C_2 \times C_2$, 
\begin{align}
\begin{split}
&X_{10}(7)=i,\quad X_{10}(19)=1,\quad X_{10}(11)= -1,\\
&X_{12}(7)=-i,\quad X_{12}(19)= 1,\quad X_{12}(11)= -1 \,. 
\end{split}
\end{align}

\subsection{Relation to Solutions of MLDE} \label{subsec:mlderel}

Suppose we have a set of RCFT characters
\be
 \chi_i= q^{\alpha_i}\sum_{n=0}^\infty c_i(n) q^n , ~~~~i=1, 2, \cdots n \, ,
 \ee
that arise as the solution of an $n$th order MLDE such
that the modular Wronskian has $\ell=0$ as was the case for the minimal and affine models discussed earlier. If the conductor of the RCFT is $N$ then we can write
$\alpha_i= m_i/N$ for some $m_i \in \IZ$. The Hecke images $\mathsf{T}_p$ are characters with the same conductor $N$ but with new leading exponents $\alpha^{(p)}_i $ 
determined by the minimum integers $n_i$ such that $b_i^{(p)}(n_i) \ne 0$ in eqn.(\ref{HeckeRcftCoef}). 
As a result of eqn.(\ref{eqn:rroch}), the Hecke images will thus have to satisfy an MLDE of the same order, but with different parameters and with in general a different value of the parameter $\ell$
which counts the zeroes of the modular Wronskian. In other words, the action of the Hecke operators on solutions induces an action on the modular Wronskians which in general
changes the structure of the divisor of $W$. 

As an example consider the Yang-Lee model.  We then have for the leading exponents of the Hecke image 
\be
\alpha^{(p)}_0 = - \frac{p}{60}, \quad
\alpha^{(p)}_1 = \frac{(11p) ~{\rm mod~} 60}{60} \, .
\ee
We denote the corresponding $\ell$ values computed from eqn.(\ref{eqn:rroch})  by  $\ell^{YL}(p)$ and provide a list of values for small $p$ in Table \ref{table:ellyl}. 
This suggests that the Hecke images of Yang-Lee characters with $p=7, 13, 19$ might already be included in the list of solutions to the $n=2$ MLDE with $\ell=0$ and indeed
we will find this to be the case. Similarly we will identify the Hecke images with $p=41, 47, 53, 59$ with some of the entries in Table 1 of \cite{Hampapura:2015cea}. The Hecke
images with $p=11,17,23,29$ do not appear in \cite{Hampapura:2015cea} because the characters contain negative integer coefficients for reasons that will be explained in the following section. Note also that no Hecke images of the YL model appear with $2< \ell < 6$ which is consistent with the
result found in \cite{Hampapura:2015cea} on the absence of solutions with these $\ell$ values.

Another important example is the Ising model. In this case the leading exponents of the Hecke images are
\begin{align}
\alpha^{(p)}_0 = - \frac{p}{48},\quad
\alpha^{(p)}_1 = \frac{(23p) ~{\rm mod~} 48}{48}, \quad
\alpha^{(p)}_2 = \frac{(2p) ~{\rm mod~} 48}{48} \, .
\end{align}
Denote the corresponding $\ell$ values by $\ell^{I}(p)$, which are listed in Table \ref{table:ellIsing}.
Unlike the Hecke images of Yang-Lee characters, all $\ell^I(p)$ equal $0$ when $0<p<48$. This ensures that there are no zeros in the modular Wronskian when the Hecke images with $p<48$ are viewed as solutions to a third order MLDE.  Increasing $p$ by a multiple of $48$ increases $\ell^I(p)$ by the same multiple of $6$.
Note that the Hecke image with $p=47$ coincides with the Baby Monster characters obtained from MLDE in \cite{Hampapura:2016mmz}.

\begin{center} 
\begin{table}
\begin{tabular}{c|cccccccccccccccccc}
\hline
$p$ & 7 & 11 & 13 & 17 & 19 & 23 & 29 & 31 & 37 & 41 & 43 & 47& 49 & 53 & 59 & 61 & 67 & 71 \\
$\ell^{YL}(p)$ & 0 & 2 & 0 & 2 & 0  & 2 &2 & 0 & 0 & 2 & 0 & 2 & 0 & 2 &2  & 6 & 6 & 8  \\
\hline
\end{tabular}
\caption{Number of zeros in the modular Wronskian for Hecke images under $\mathsf{T}_p$ of Yang-Lee characters for small values of $p$.} \label{table:ellyl}
\end{table}
\end{center}

\begin{center} 
\begin{table}
\begin{tabular}{c|ccccccccccccccccccc}
\hline
$p$ & 5 & 7 & 11 & 13 & 17 & 19 & 23 & 25 & 29 & 31 & 35 & 37 & 41 & 43 & 47 & 49 & 53 & 55   \\
$\ell^{I}(p)$ & 0 & 0 & 0 & 0 & 0  & 0 & 0 & 0 & 0 & 0 & 0 & 0 & 0 & 0 & 0 & 6 & 6 & 6   \\
\hline
\end{tabular}
\caption{Number of zeros in the modular Wronskian for Hecke images under $\mathsf{T}_p$ of Ising characters for small values of $p$.} \label{table:ellIsing}
\end{table}
\end{center}

\section{Applications and Examples}\label{sec:Applications}

\subsection{One Character RCFT}

RCFT with a single character can only exist for $c=0 ~{\rm mod}~8$ with the simplest example being affine $E_8$ at level one with
character
\be
\chi^{E_8} = \frac{E_4}{\eta^8} = \frac{\Theta_{E_8}}{\eta^8}
\ee
with $E_4$ the weight four Eisenstein series (and the theta function of the $E_8$ root lattice) and $\eta$ the Dedekind eta function.
This is a somewhat degenerate case of our formalism, but nonetheless Hecke operators still act in an interesting way. $\chi^{E_8}$ transforms
according to a one-dimensional representation of the modular group with $\rho(S)=1$ and $\rho(T)=e^{2 \pi i/3}$. It thus has conductor $3$.
By comparing $q$ expansions one deduces that
\begin{align}
\mathsf{T}_2 \chi^{E_8} &= \chi^{E_8 \times E_8}= \chi^{Spin(32)/\IZ_2} = \frac{E_4^2}{\eta^{16}} \\
\mathsf{T}_4 \chi^{E_8} &= \chi^{{\rm ext},32} = q^{-4/3} \left( 1+139504 q^2 + 69332992 q^3 + \cdots \right) \\
\mathsf{T}_5  \chi^{E_8} &= \chi^{{\rm ext},40} = q^{-5/3} \left( 1+20620 q^2 + 86666240 q^3 + \cdots \right)
\end{align}
where   $\chi^{E_8 \times E_8}$ and $\chi^{Spin(32)/\IZ_2}$ are the characters of lattice VOAs associated to the two rank $16$ even, self-dual lattices and  $\chi^{{\rm ext},32}$ and $\chi^{{\rm ext},40}$ are the characters of extremal  VOAs of rank $32$ and $40$ appearing in Table 5.1
of \cite{hoehn}.  Closely related results were obtained in \cite{Witten:2007kt} by looking at the Hecke images of CFT with a single character
at $c=24$ where the usual Hecke operators for $SL(2,\IZ)$ can be used and again the images correspond to certain extremal CFT.

Hecke operators $\mathsf{T}_p$ for $p \ge 7$ acting on the affine $E_8$ character lead to modular forms with a gap of more
than $2$ between the powers of $q$ of the leading and subleading terms with nonzero coefficients and thus cannot be characters of a RCFT
since the state $L_{-2} |0 \rangle$ is always present in an RCFT.  This can be remedied using the same idea as in \cite{Witten:2007kt} by adding
together Hecke images $\mathsf{T}_{p'} \chi^{E_8}$ for $p'<p$, $p'=0 ~{\rm mod~} 3$ in a way that leads to a character consistent with the existence of a decomposition into Virasoro characters.

\subsection{Two Character RCFT}

We now apply the Hecke operators of the previous section to derive relations between the characters of  RCFTs with more than
one character. These examples in general involve mixing between the different components of the characters. We will start with the Yang-Lee model.
The characters were given in eqns. (\ref{ylone}, \ref{yltwo}). The representations of $S$ and $T$ are
\be
\rho^{YL}(S) = \frac{2}{\sqrt{5}} 
\begin{pmatrix}  ~\sin(\frac{2 \pi}{5}) & ~\sin(\frac{\pi}{5}) \\  \sin(\frac{\pi}{5}) & -\sin(\frac{2 \pi}{5})  \end{pmatrix}, \qquad \rho^{YL}(T) = {\rm diag}( \xi_{60}^{-1}, \xi_{60}^{11}),
\ee
where we use the notation $\xi_N$ for $e^{2 \pi i/N}$. We thus see that the conductor is $N=60$. 

We now consider which $\mathsf{T}_p$ acting on the characters $\chi^{YL}_i$ can lead to consistent RCFT characters by which we mean that the vacuum has degeneracy one,
the coefficients in the $q$ expansion are non-negative integers and the Verlinde formula for the representation $\rho(\sigma_p)$ leads to non-negative fusion coefficients.  Using
eqns. (\ref{ylsigp}) and (\ref{HeckeRcftCoef}) we can rule out $p=11,29,31,49 {\rm ~mod~} 60$ since these lead to a negative coefficient $b_0^{(p)}(-p)= \rho_{00}(\sigma_p) b_0^{YL}(-1) = -1$. 
Similarly we can rule out $p= 17, 23, 37, 43 {\rm ~mod~} 60$ since then $b_1^{(p)}(-p)= \rho_{10}(\sigma_p) b_0(-1)= -1$. This leaves $p=19,41, 59 {\rm ~mod~} 60$ with $\rho^{YL}(\sigma_p)$
the identity matrix so that the Hecke image has no mixing between different Yang-Lee characters and $p=7,13,47, 53 {\rm ~mod~}60$ with off-diagonal $\rho^{YL}(\sigma_p)$ and hence
mixing between the two characters in the Hecke image. Of these values of $p$ only $p=7,13,19$ have $\ell=0$ according to Table \ref{table:ellyl}.

For $p=7,13$ we have
\be
\rho^{YL}(\sigma_{7})=\rho^{YL}(\sigma_{13})=\begin{pmatrix}
0 & -1\\ 1 &0
\end{pmatrix} \, .
\ee
We can then use eqn.(\ref{HeckeRcftCoef}) to derive the coefficients $b^{(p)}_i(n)$ of the Hecke image. Although the Hecke relations between coefficients are most easily seen when characters
are written as a series in $q^{1/N}$, it is more conventional in conformal field theory to write characters in terms of a leading fractional power of $q$ and a series with integer powers of $q$.
Thus for the Yang-Lee model we define coefficients of characters as
\begin{align}
\chi^{YL}_0 = q^{-1/60} \sum_{n=0}^\infty c^{YL}_0(n) q^n \, ,\\
\chi^{YL}_{1/5} = q^{11/60} \sum_{n=0}^\infty c^{YL}_{1/5}(n) q^n \, , \nonumber
\end{align}
and we similarly define coefficients $c^{G_2}_0(n)$ and $c^{G_2}_{2/5}(n)$ of the vacuum character and character corresponding to the primary of conformal weight $2/5$
for level one affine $G_2$ and coefficients $c^{F_4}_0(n)$ and $c^{F_4}_{3/5}(n)$ for level one affine $F_4$ characters.

We now show that the modular representation and characters of affine level one $G_2$ and $F_4$ are the Hecke images of the Yang-Lee characters under
the Hecke operators $\mathsf{T}_7$ and $\mathsf{T}_{13}$ respectively.  The relation between modular representations is well known. We have
\begin{align}
\begin{split}
&\rho^{G_2}(S)=\rho^{F_4}(S)=\rho^{YL}(\sigma_p S)
=\rho^{YL}(\sigma_p) \rho^{YL}(S)\\ 
=&
\begin{pmatrix}
0 & -1 \\1 &0
\end{pmatrix} \cdot
\frac{2}{\sqrt{5}}
\begin{pmatrix}
      \sin \left(\frac{2\pi}{5}\right) &
      \sin \left(\frac{ \pi}{5}\right) 
      \\
      \sin \left(\frac{\pi}{5}\right) &
      -    \sin \left(\frac{2\pi}{5}\right) 
    \end{pmatrix}
    =\frac{2}{\sqrt{5}}
\begin{pmatrix}
     - \sin \left(\frac{\pi}{5}\right) &
      \sin \left(\frac{ 2\pi}{5}\right) 
      \\
      \sin \left(\frac{2\pi}{5}\right) &
       \sin \left(\frac{\pi}{5}\right) 
    \end{pmatrix}.
\end{split}
\end{align}
and
\begin{align}
\rho^{G_2}(T)  &= \rho^{YL}(T^{\bar 7})=\rho^{YL}(T^{43})= {\rm diag}\left(\xi_{60}^{17},~ \xi_{60}^{-7} \right), \\
\rho^{F_4}(T) &= \rho^{YL}(T^{\bar{13}}) = \rho^{YL}(T^{37}) = {\rm diag}\left(\xi_{60}^{23}, ~\xi_{60}^{-13} \right) .\nonumber
\end{align}
Note that the ordering of the $G_2$ and $F_4$ characters comes out with the vacuum character in the second entry due to our choice of ordering of the
Yang-Lee characters. 

Translating the Hecke relation into a relation between coefficients in the $q$ expansion above we find that the Hecke relations $\chi^{G_2} =\mathsf{T}_7 \chi^{YL}$
and $\chi^{F_4}= \mathsf{T}_{13} \chi^{YL}$ are equivalent to the relations
\begin{align} \label{G2 Hecke}
\begin{split}
c_0^{G_2} (n)=&
\begin{cases} 7c_{1/5}^{YL} (7n-1)
& \text{if  } 7 \nmid n , \\
7c_{1/5}^{YL} (7n-1) + c_0^{YL} (\frac{n}{7})
&  \text{if  } 7 | n ;  
\end{cases} \\
c_{2/5}^{G_2} (n)=&
\begin{cases} 7c_{0}^{YL} (7n+2)
& \text{if  } 7 \nmid (n-1) , \\
7c_{0}^{YL} (7n+2) - c_{1/5}^{YL} (\frac{n-1}{7})
&  \text{if  } 7 | (n-1) . 
\end{cases} 
\end{split}
\end{align}
\begin{align} \label{F4 Hecke}
\begin{split}
c_0^{F_4} (n)=&
\begin{cases} 13c_{1/5}^{YL} (13n-3)
& \text{if  } 13 \nmid n , \\
13c_{1/5}^{YL} (13n-3) + c_0^{YL} (\frac{n}{13})
&  \text{if  } 13 | n ;  
\end{cases} \\
c_{3/5}^{F_4} (n)=&
\begin{cases} 13c_{0}^{YL} (13n+5)
& \text{if  } 13 \nmid (n-2) , \\
13c_{0}^{YL} (13n+5) - c_{1/5}^{YL} (\frac{n-2}{13})
&  \text{if  } 13 | (n-2) .
\end{cases} 
\end{split}
\end{align}
which we claim are true.

We should clarify what we mean when we write down relations like these and others to be presented later. From a practical point of view,
sufficient for most physicists, we have verified these relations  to high order in their $q$ expansions, typically  to order $q^{2000}$. This
can be turned into a precise mathematical statement although we have not done this for all cases. We have done it for the example
above as follows.  

The Sturm bound \cite{sturm} states that two modular forms  of weight $k$ for a congruence subgroup $\Gamma$ of $SL(2,\IZ)$ are equal if
the first $ k[SL(2,\IZ):\Gamma]/12$ coefficients in their $q$ expansion agree where $[SL(2,\IZ):\Gamma]$ is the index of $\Gamma$ in
$SL(2,\IZ)$. We can apply this by noting that
$\eta (\tau)^2$ is a weight one modular form for $\Gamma(12)$ so we can multiply each term of the putative identity which involves modular forms on  $\Gamma(60)$ by $\eta^2$ to obtain weight one forms for $\Gamma(60)$.  The index of $\Gamma(60)$ in $SL(2,\IZ)$ is $138240$ so the Sturm bound implies that the two modular forms are equal if their $q$ expansions match to order $q^{11520}$.  We have verified the above identities to this
order thus proving that the Hecke images of YL characters under $\mathsf{T}_7$ and $\mathsf{T}_{13}$ are the affine level one $G_2$ and $F_4$ characters as claimed.

The same analysis with the obvious changes identifies the characters of the $E_{7 \frac{1}{2}}$ model with the Hecke image of the Yang-Lee characters under $\mathsf{T}_{19}$:

\begin{align} \label{E7_1/2 Hecke}
\begin{split}
c_0^{E_{7+1/2}} (n)=&
\begin{cases} 19 c_{0}^{YL} (19n-6)
& \text{if  } 19 \nmid n , \\
19 c_{0}^{YL} (19n-6) + c_0^{YL} (\frac{n}{19})
&  \text{if  } 19 | n ;  
\end{cases} \\
c_{4/5}^{E_{7+1/2}} (n)=&
\begin{cases} 19 c_{1/5}^{YL} (19n+9)
& \text{if  } 19 \nmid (n-3) , \\
19c_{1/5}^{YL} (19n+9) + c_{1/5}^{YL} (\frac{n-3}{19})
&  \text{if  } 19 | (n-3) .  
\end{cases} 
\end{split}
\end{align}
There is no character mixing here, since
\begin{align}
\rho^{YL}(\sigma_p)=\mathbb{I}_2  ~,
\qquad p=19.
\end{align}

For $p=41, 47, 53, 59$ which are the remaining $p<60$ which lead to Hecke images with positive coefficients the Hecke images have $\ell=2$ and
can be identified with entries Nos. 1,4,6,9 in Table 1 of \cite{Hampapura:2015cea}. Since the representation matrix $\rho^{YL}(\sigma_p S)$ for the theories
with $p=41, 59$ are the same as those of the $E_{7 \frac{1}{2}}$ theory the resulting theory will have some negative integer fusion coefficients and thus the characters cannot
be characters of a consistent RCFT but perhaps can be identified with characters of Intermediate Vertex subalgebras as was done for the $E_{7 \frac{1}{2}}$ theory
in \cite{kawasetsu}. 

While $\rho^{YL}(\sigma_p)$ is periodic with period $N=60$, the Hecke operators $\mathsf{T}_p$ are not. The Hecke images $\mathsf{T}_p \chi^{YL}$ for 
$p=7,13, 47, 53 ~{\rm mod~} 60$ by a small extension of the above discussion can be shown to have all positive integer coefficients in their $q$ expansion, have degeneracy one for the vacuum representation and have modular S matrix $\rho^{YL}(\sigma_p S)$ which is the same as that for the affine $G_2$ and $F_4$ theories which are known to lead to non-negative integer fusion coefficients.  However, as in the one character case these characters are not consistent with a decomposition into Virasoro characters for sufficiently large $p$ but again this can be remedied by combining
together different Hecke images.  If we take $P=60M+p$ for $p=7,13, 47, 53 ~{\rm mod~} 60$ then the component of the Hecke image of the Yang-Lee characters we would
like to identify with the vacuum characters has a $q$ expansion of the form
\be
q^{-p/60}(q^{-M}+O(q)) \, .
\ee
In a RCFT this $q$ expansion must contain at least the Virasoro vacuum character at $c=2*P/5$ which is
\be
{\rm ch}_{c=2P/5,h=0}(q)=\frac{q^{-P/60}}{\prod_{n=2}^\infty (1-q^n)}= q^{-P/60} \sum_{n=0}^\infty c(n) q^n
\ee
The combination of Hecke images
\be
\mathsf{T}_{60M+p} \chi^{YL}+ \sum_{k=0}^{M-2} d(k) \mathsf{T}_{p+60*k} \chi^{YL}
\ee
will thus have a decomposition into a Virasoro vacuum character plus Virasoro characters at larger conformal dimension $h$ provided
that $d(k) \ge c(k)$. The case $d(k)=c(k)$ might be regarded as a two-character generalization of the  one character extremal CFTs explored in
\cite{hoehn, Witten:2007kt} and later references. 
These Hecke images thus provide a completely consistent set of possible RCFT characters.
It would clearly be interesting to see if these can be identified with characters of specific RCFTs.

There are other characters related by Hecke relations in the Deligne-Cvitanovi{\'c} exceptional series.
For example, the $A_1$ and $E_7$ affine characters at level $k=1$  are characters of RCFTs with conductor $N=24$ and the Fourier coefficients of the $E_7$ characters are Hecke images
under $\mathsf{T_7}$ of those of $A_1$:
\begin{align} \label{A1_E7 Hecke}
\begin{split}
c_0^{E_7} (n)=&
\begin{cases} 7c_{0}^{A_1} (7n-2)
& \text{if  } 7 \nmid n , \\
7c_{0}^{A_1} (7n-2) + c_0^{A_1} (\frac{n}{7})
&  \text{if  } 7 | n ;  
\end{cases} \\
c_{3/4}^{E_7} (n)=&
\begin{cases} 7c_{1/4}^{A_1} (7n+3)
& \text{if  } 7 \nmid (n-1) , \\
7c_{1/4}^{A_1} (7n+3) + c_{1/4}^{A_1} (\frac{n-1}{7})
&  \text{if  } 7 | (n-1) . 
\end{cases} 
\end{split}
\end{align}
where we have used
\begin{align}
\rho^{A_1}(\sigma_7)=\mathbb{I}_2~.
\end{align}

It also turns out that four more of the entries in Table 1 of \cite{Hampapura:2015cea} can also be constructed as Hecke images of $n=2, \ell=0$ models.
In addition to the examples mentioned earlier, the $c=17,23$ examples are Hecke images under $\mathsf{T}_{17}$ and $\mathsf{T}_{23}$  respectively of the affine $A_1$ characters and the $c=20,22$ examples are Hecke images under $\mathsf{T}_{5}$ of the $D_4$ characters and under
$\mathsf{T}_{11}$ of the $A_2$ characters respectively.

The only entries in the Deligne exceptional series we have not related by Hecke operators are affine $E_6$ and affine $A_2$. Their
characters are related, but the relationship is more complicated, presumably because they involve a Hecke operator for the prime $3$
which is not relatively prime to the conductor $12$. We presume these characters can also be related to the remaining $c=18$ example with $n=2, \ell=2$
but the details remain to be worked out.

\subsection{Three Character RCFT}

Hecke relations between characters of RCFTs appear to be quite common. Here we explore a few examples of relations between RCFTs with three independent characters. Certainly the most well known RCFT with three independent characters is the Ising model with $c=1/2$, equivalent to
a free Majorana fermion. 

The well known characters and representation matrices of the Ising model show that the conductor is $N=48$.  Denoting the Ising model modular
representation matrices by $\rho^I$, there are
only two different values of $\rho^I(\sigma_p)$ that occur depending on whether $p^2=1 {\rm ~mod~} 48$ or $p^2= 25 {\rm ~mod~} 48$:
\begin{align}
\rho^I(\sigma_p)=
\mathbb{I}_3~,
\qquad p^2=1 \text{ mod }48~.
\end{align}
\begin{align}
\rho^I(\sigma_p)=
\left(
\begin{array}{ccc}
 0 & 1 & 0 \\
 1 & 0 & 0 \\
 0 & 0 & -1 \\
\end{array}
\right)~,
\qquad p^2= 25 \text{ mod }48~.
\end{align}

The Hecke images $\mathsf{T}_p \chi^I$ for all prime $p$ with $(p,48)=1$ lead to characters obeying all of the consistency conditions for
RCFT characters. We mention here some examples where we have been able to match the Hecke images to the characters of known RCFT.
For all $p<24$ the Hecke images are the characters of  affine $so(p)$ algebras at level 1. 
While for all $24<p<48$, $\mathsf{T}_p\chi^I$ is a shift from the $\hat{so}(p)_1$ characters :
\begin{align}
\begin{pmatrix}
\chi_0 \\ \chi_{v} \\ \chi_{s}
\end{pmatrix}^{so(p)}
-
\begin{pmatrix}
\chi_0 \\ \chi_{v} \\ \chi_{s}
\end{pmatrix}^{Y=(p)}
=p \,\CP \cdot
\begin{pmatrix}
\chi_0 \\ \chi_{v} \\ \chi_{s}
\end{pmatrix}^{so(p-24)}   ,
\end{align}
where
\begin{align}
\CP=\left(
\begin{array}{ccc}
 0 & 1 & 0 \\
 1 & 0 & 0 \\
 0 & 0 & -1 \\
\end{array}
\right)~.
\end{align}
The super index $Y=(p)$ represents that it is the Hecke image under $\mathsf{T}_p$.
The subindices $0,v,s$ of $\chi$ stand for the vacuum, vector and spinor representation respectively.

For $p=31$ the Hecke image of the Ising model characters are the characters of affine $E_8$ at level $2$:
\begin{align}
\begin{split}
\chi^{E_8^{(2)}}_0 =& q^{-31/48}(1+248 q+ 31124 q^2+ 871627 q^3+ 13496501 q^4 +\cdots),\\
\chi^{E_8^{(2)}}_{3/2} =& q^{41/48}(3875+181753 q+ 3623869 q^2+ 46070247 q^3+\cdots),\\
\chi^{E_8^{(2)}}_{15/16} =& q^{7/24}(248+34504 q+ 1022752 q^2+ 16275496 q^3+\cdots).
\end{split}
\end{align}
with the Hecke relations to the Ising model characters given explicitly by
\begin{align} \label{p=31 Hecke}
\begin{split}
c_0^{E_8^{(2)}} (n)=&
\begin{cases} 31c_0^I (31n-20)
& \text{if  } 31 \nmid n , \\
31c_0^I (31n-20)+ c_0^I (\frac{n}{31})
&  \text{if  } 31 | n ;  
\end{cases} \\
c_{3/2}^{E_8^{(2)}} (n)=&
\begin{cases} 31c_{1/2}^I (31n+26)
& \text{if  } 31 \nmid (n-14) , \\
31c_{1/2}^I (31n+26)+ c_{1/2}^I (\frac{n-14}{31})
&  \text{if  } 31 | (n-14) ;  
\end{cases} \\
c_{15/16}^{E_8^{(2)}} (n)=&
\begin{cases} 31c_{1/16}^I (31n+9)
& \text{if  } 31 \nmid (n-1) , \\
31c_{1/16}^I (31n+9)+ c_{1/16}^I (\frac{n-1}{31})
&  \text{if  } 31 | (n-1) .
\end{cases} 
\end{split}
\end{align}

For $p=47$ case the Hecke images are the characters of the Baby Monster vertex algebra constructed in \cite{hoehn}.
The Baby Monster characters are
\begin{align}
\begin{split}
\chi^{BM}_0 =& q^{-47/48}(1+ 96256 q^2+ 9646891 q^3+ 366845011 q^4 +\cdots),\\
\chi^{BM}_{3/2} =& q^{25/48}(4371+1143745 q + 64680601 q^2 + 1829005611 q^3+\cdots),\\
\chi^{BM}_{31/16} =& q^{23/24}(96256+10602496 q + 420831232 q^2 + 9685952512 q^3+\cdots).
\end{split}
\end{align}
and are given in terms of Ising model characters by
\begin{align} \label{p=47 Hecke}
\begin{split}
c_0^{BM} (n)=&
\begin{cases} 47c_0^I (47n-46)
& \text{if  } 47 \nmid n , \\
47c_0^I (47n-46)+ c_0^I (\frac{n}{47})
&  \text{if  } 47 | n ;  
\end{cases} \\
c_{3/2}^{BM} (n)=&
\begin{cases} 47c_{1/2}^I (47n+24)
& \text{if  } 47 \nmid (n-22) , \\
41c_{1/2}^I (47n+24)+ c_{1/2}^I (\frac{n-22}{47})
&  \text{if  } 47 | (n-22) ;  
\end{cases} \\
c_{31/16}^{BM} (n)=&
\begin{cases} 47c_{1/16}^I (47n+45)
& \text{if  } 47 \nmid (n-1) , \\
47c_{1/16}^I (47n+45) + c_{1/16}^I (\frac{n-1}{47})
&  \text{if  } 47 | (n-1) ;  
\end{cases} 
\end{split}
\end{align}

We now give a few example of Hecke images of the minimal model $\mathsf{M}(2,7)$ characters. 
The representations of $S$ and $T$ are
\begin{align}
\rho^{\mathsf{M}(2,7)}(S) &= \frac{2}{\sqrt{7}} \begin{pmatrix}  
\sin(\frac{4 \pi}{7}) & \sin(\frac{2 \pi}{7}) & \sin(\frac{\pi}{7}) \\  
 \sin(\frac{2 \pi}{7}) & -\sin(\frac{\pi}{7}) & -\sin(\frac{4 \pi}{7}) \\
 \sin(\frac{\pi}{7}) & -\sin(\frac{4 \pi}{7}) & \sin(\frac{2 \pi}{7})
 \end{pmatrix}, \nonumber \\
\qquad \rho^{\mathsf{M}(2,7)}(T) &= {\rm diag}( \xi_{42}^{-1}, \xi_{42}^{5},\xi_{42}^{17}).
\end{align}

For integers $p$ with $\text{gcd}(p,42)=1$ we find
\begin{align}
\rho^{\mathsf{M}(2,7)}(\sigma_p)=
\mathbb{I}_3~,
\qquad p^2=1 \text{ mod }42~.
\end{align}
\begin{align}
\rho^{\mathsf{M}(2,7)}(\sigma_p)=
\left(
\begin{array}{ccc}
 0 & 1 & 0 \\
 0 & 0 & -1 \\ 
 -1 & 0 & 0 \\
\end{array}
\right)~,
\qquad p^2=25 \text{ mod }42~.
\end{align}
\begin{align}
\rho^{\mathsf{M}(2,7)}(\sigma_p)=
\left(
\begin{array}{ccc}
 0 & 0 & -1 \\
 1 & 0 & 0 \\ 
 0 & -1 & 0 \\
\end{array}
\right)~,
\qquad p^2=37 \text{ mod }42~.
\end{align}

The Hecke image under $p=5$ gives rise to a vector-valued modular form that is relevant to the WRT invariant of the Brieskorn homology sphere $\Sigma(2,3,7)$ \cite{Hikami:Spherical_Seifert}.
The Hecke image under $p=59$  recovers the three character model without Kac-Moody symmetry
\begin{align}
\begin{split}
\chi_0^{(59)} &=q^{-59/42}
(1 + 63366 q^2 + 46421200 q^3 + 5765081101 q^4 + \cdots ), \\
\chi_{16/7}^{(59)} &=q^{37/42}
(715139 + 257698784 q + 24078730130 q^2 + \cdots ), \\
\chi_{17/7}^{(59)} &=q^{43/42}
( 848656 + 232637826 q+ 19201964416 q^2 + \cdots ),
\end{split}
\end{align}
obtained in \cite{Hampapura:2016mmz} by analyzing the third order MLDE.

\subsection{Bilinear Relations between Hecke Images and Modular Functions}

In \cite{Hampapura:2016mmz} some bilinear relations between RCFT characters  and the modular $J$ function were exhibited. 
These turn out to follow simply from properties of characters and their Hecke images. 
Let $f_1(\tau)$ and $f_2(\tau)$ be the Hecke images of some 
RCFT character $\chi(\tau)$:
\begin{align}
f_i (\tau) = (\mathsf{T}_{p_i} \chi )(\tau),
\qquad i=1,2.
\end{align}
The induced modular representations are
\begin{align}
\rho^{(p_i)}(T)= \rho(T^{\bar p_i}) ,
\qquad 
\rho^{(p_i)}(S)= G_{p_i}^{-1} \rho(S)= \rho(S) G_{p_i} \, .
\end{align}

Under the action of $SL(2,\IZ)$ generators $T$ and $S$, the bilinear form $f_2(\tau)^T G_{\ell} f_1(\tau)$ transforms as
\begin{align}
\begin{split}
f_2(\tau+1)^T G_{\ell} f_1(\tau+1) &= 
f_2(\tau)^T \big[ \rho(T^{\bar p_2}) G_{\ell} \rho(T^{\bar p_1}) \big] f_1(\tau) , \\
f_2(-1/\tau)^T G_{\ell} f_1(-1/\tau) &= 
f_2(\tau)^T \big[ G_{p_2}^{-1}\rho(S) G_{\ell} \rho(S) G_{p_1} \big] f_1(\tau) \, .
\end{split}
\end{align}
and will be modular invariant provided that
\begin{align} \label{condition:ModularInvariance}
\begin{split}
 \rho(T^{\bar p_2}) G_{\ell} \rho(T^{\bar p_1}) =& G_{\ell}, \\
 G_{p_2}^{-1}\rho(S) G_{\ell} \rho(S) G_{p_1} =& G_{\ell}  \, ,
 \end{split}
\end{align}
or equivalently
\begin{align}
\rho\big( T^{\bar p_2+\bar p_1 \ell^2} \big) =& \mathbb{I}_n,
\label{T:invariance2} \\
 G_{p_2}^{-1} G_{-\bar \ell} G_{p_1}G_{\bar \ell} 
 =& \mathbb{I}_n. \label{S:invariance2}
\end{align}
Since the set of $G_{\ell}$ is homomorphic to $(\IZ/N\IZ)^{\times}$ and hence abelian, eqn.\eqref{S:invariance2} amounts to 
\begin{align}
G_{-p_1\bar p_2 \bar\ell^2} = \mathbb{I}_n.
\end{align}
While eqn.\eqref{T:invariance2} states the fact
\begin{align}\label{T:invariance3}
\bar p_2 +\bar p_1 \ell^2 \equiv 0 \text{ mod }N,
\end{align}
which necessarily implies $-p_1\bar p_2 \bar\ell^2 \equiv 1 \text{ mod }N$ therefore solves eqn.\eqref{S:invariance2}.

As an application consider the bilinear combination presented in \cite{Hampapura:2016mmz} 
of characters of $\mathsf{M}(2,7)$ and its dual $\widetilde{\mathsf{M}}(2,7)$. 
The $\widetilde{\mathsf{M}}(2,7)$ characters are the Hecke images of the $\mathsf{M}(2,7)$ characters under $\mathsf{T}_{59}$. A modular invariant bilinear form
can be constructed by taking
\begin{align}
p_1=1,\quad  p_2 =59, \quad
\ell^2 = -17  \text{ mod } 42,
\end{align}
which satisfy the criteria eqn.\eqref{T:invariance3}.
The leading coefficient is easily seen to be $q^{-1}$ so this combination must be the modular $J$ function
\begin{align}
J(q)=\frac{1}{q}+196884 q + 21493760 q^2 + 864299970 q^3 +\cdots
\end{align}
up to an additive constant.

For the Ising model with conductor $N$ equals $48$ one can also construct modular invariant combinations. For example,
when $\ell$ is taken to be 1, eqn.\eqref{T:invariance3} boils down to
\begin{align}
p_1+p_2 = 0 \text{ mod }48.
\end{align}
As a result, $ (\mathsf{T}_{p} \chi )^T(\mathsf{T}_{48-p} \chi )$ is  modular invariant for $p\in (\IZ/48\IZ)^{\times}$.
Since it has the most singular term $q^{-1}$, this bilinear form also gives $J(q)$ up to an additive constant, namely
\begin{align}
 (\mathsf{T}_{p} \chi )^T(\mathsf{T}_{48-p} \chi ) 
 =J(q)+c_0^{(p)}(1)+c_0^{(48-p)}(1) .
\end{align}

\section{Conclusions and Outlook}\label{sec:Conclusions}

We have established that the known Galois symmetry connecting modular representations of certain RCFTs has an extension to a relation via Hecke operators between their characters. We explored these connection for theories with two or three independent characters and showed that many of the characters are Hecke images of the characters of minimal models, their tensor products or characters
of affine Lie algebras. We also showed that Hecke operators connect solutions to MLDE which have different numbers of zeroes in the modular
Wronskian and used this to show that all the $\ell=2$ characters found in Table 1 of \cite{Hampapura:2015cea} are Hecke images of $\ell=0$ solutions. 
 We also showed that some earlier observations regarding the pairing of characters of different RCFTs into the modular $J$ function have a
simple interpretation in terms of Hecke relations between characters. In general it appears that given the characters $\chi_i$ of a known RCFT with $n$ independent characters
then there will be an infinite sequence of characters $(\mathsf{T}_p \chi)$ that will, for particular values of $p$ depending on $\rho(\sigma_p)$, provide characters that obey all the conditions
required for them to be characters of a RCFT, that is vacuum degeneracy one, positive integer coefficients in the $q$ expansion and non-negative integer fusion coefficients. For
the Yang-Lee model we gave an explicit description of the allowed $p$ values. This can clearly be extended to other RCFT in a straightforward way. It will be interesting to see if these
infinite sets of Hecke images can be connected with explicit RCFTs.  We would also like to achieve a more global picture of how Hecke operators relate the characters of
different RCFTs and also whether the Hecke relations extend more broadly to the full structure of the RCFT. In particular the results presented here
are compatible with the idea that the characters of all RCFT with $n$ independent characters are Hecke images of the characters of a finite set of RCFTs. In the cases examined here this finite set consists of minimal models and models with affine Lie algebras. 

From a physical point of view perhaps the most interesting question is whether there is a physical origin of the Hecke operators $\mathsf{T}_p$. Hecke operators have appeared previously in the computation of the elliptic genus of symmetric products of K{\"a}hler manifolds where they do have a clear physical interpretation in terms of instanton sectors but that interpretation does not apply in any obvious way here.  Characters of vertex operator algebras have recently appeared in the context of the Schur index of $N=2$ superconformal theories in four dimensions \cite{Beem} and this seems like an interesting context in which to look for applications of the Hecke relations presented here. If the Hecke images of characters $\mathsf{T}_p \chi$ that satisfy all the consistency conditions for characters of a RCFT actually are characters of RCFT then at large $p$ the central charge becomes large and it is then natural to ask about a dual $AdS_3$ description of the Hecke images.

\section*{Acknowledgements}

We would like to thank A. Blass, P. Bantay, F. Brunault, J. Chi, T. Creutzig, M. Emerton, M. Kaneko, M. Kim, G. Moore, K. Ono, M. Raum and J. Rouse for helpful discussions and correspondence.  
J.H. and Y.W. acknowledge support from the NSF \footnote{Any opinions, findings, and conclusions or recommendations expressed in this material are those of the author(s) and do not necessarily reflect the views of the National Science Foundation.} under grant PHY 1520748. 

\appendix

\section{Properties of  the Hecke operator $\mathsf{T}_p$} \label{app:Hecke}
The properties eqn.\eqref{Tm_Tn} and \eqref{T_p^n} are special cases of the general identity
\begin{align}\label{T_general}
T_m T_n=\sum_{d|\text{gcd}(m,n)} d^{k-1} T_{mn/d^2}~,
\end{align}
for the standard Hecke operators acting on modular forms of weight $k$.

In this section we provide the analog of eqn.\eqref{T_general},
based on which the counterparts of eqn.\eqref{Tm_Tn} and \eqref{T_p^n} can be found as particular cases.
Namely the
Hecke operators $\mathsf{T}_m$, $\mathsf{T}_n$ for $m,n$ relatively prime obey
\begin{align} \label{Tm_Tn:new}
\mathsf{T}_m \mathsf{T}_n f = \mathsf{T}_{mn} f
\end{align}
and for $n=p^{m}$ with $p$ prime and $m \ge 1$
\begin{align} \label{T_p^n:new}
\mathsf{T}_{p^{m+1}} f= \mathsf{T}_p \mathsf{T}_{p^m} f - p^{1-k} \sigma_p \circ \mathsf{T}_{p^{m-1}} f \, .
\end{align}
The form of the equation above slightly differs from eqn.\eqref{T_p^n} due to the normalization of the Hecke operator $ \mathsf{T}_p$ for
RCFT characters.

For an positive integer $n$ with $\text{gcd}(n,N)=1$, we denote by $J_n$ the diagonal matrix 
\begin{align}
\begin{pmatrix}
1 & 0 \\ 0 & n
\end{pmatrix}
\end{align}
and define
\begin{align}
[\![ J_n ]\!] := &
\left\{
\begin{pmatrix}
a & b \\ c & d
\end{pmatrix}
\Bigg| ~
\begin{pmatrix}
a & b \\ c & d
\end{pmatrix}
\equiv J_n \text{ mod }N,~
\det \begin{pmatrix}
a & b \\ c & d
\end{pmatrix} = \det J_n
\right\},\\
(J_n) : = & \Gamma(N) J_n \Gamma(N).
\end{align}
It is proved in \cite{Rankin} that $[\![ J_n ]\!] $ and $(J_n)$ obey
\begin{align}
[\![ J_n ]\!] =& \Gamma(N) \cdot \Delta^{(n)}_N , \\
[\![ J_n ]\!] =& \sum_{q^2|n} (q\,\sigma_q) (J_{n/q^2}).
\end{align}
The Hecke operator $\mathsf{T}_n$ on $\Gamma(N)$ modular forms $f$ is defined as
\begin{align}
(\mathsf{T}_n\, f) (\tau)= n^{-k/2} \sum_{ \delta \in \Delta_N^{(n)}} (f|_k \delta)(\tau).
\end{align} 
For a composite number $n$ relatively prime to $N$, the right transversal ${\cal T}_n^+ \equiv \Delta_N^{(n)}$ reads
\begin{align}
\Delta_N^{(n)}=
\left\{  \sigma_a
\begin{pmatrix}
a & b \\ 0 & d
\end{pmatrix}
\Bigg| ~
ad=n, a>0, b=\nu N~~(0\leq \nu <d)
\right\}.
\end{align}
As a divisor of $n$, the integer $a$ can take values other than $1$ and $n$, adding to the complexity of the representatives as well as the Fourier coefficients of the Hecke images. Following eqn.\eqref{Tm_Tn:new} and \eqref{T_p^n:new}, we can write down the prime factorization of $n$ and express $\mathsf{T}_n$ in terms of polynomials of $\mathsf{T}_p$ with $p|n$. 
It can also be shown that the Hecke image $\mathsf{T}_n f$ has the modular representation  in the same way as eqn\eqref{newrep}
\begin{align} \label{newrep:n}
\rho^{(n)}(T)  = \rho(T^{\bar n}), \qquad \rho^{(n)}(S)= \rho( \sigma_n S) \, .
\end{align}

By algebraic manipulation of $[\![ J_n ]\!]$ we can deduce identities obeyed by  the Hecke operator $\mathsf{T}_n$. We move on to prove eqn\eqref{Tm_Tn:new} and \eqref{T_p^n:new}.
For coprime $m$ and $n$, the computation of double cosets proceeds as
\begin{align}\begin{split}\label{Jm_Jn}
[\![ J_n ]\!] [\![ J_m ]\!] 
= & \sum_{q^2|n,~r^2|m}
 (q\,\sigma_q) (J_{n/q^2})  (r \,\sigma_r) (J_{m/r^2}) \\
 =& \sum_{q^2|n,~r^2|m}
 (q \,\sigma_q)  (r \,\sigma_r) (J_{n/q^2})  (J_{m/r^2}) \\
  =& \sum_{q^2|n,~r^2|m}
 (qr \,\sigma_{qr}) (J_{(nm)/(qr)^2})  \\
= &[\![ J_{nm} ]\!] .
\end{split}\end{align}
The second equality comes from the commutativity \cite{Rankin}
\begin{align}
(J_n)(\sigma_p) = (\sigma_p)  (J_n),
\qquad \text{gcd}(p,N)=1.
\end{align}
Because $m$ and $n$ are coprime, $(qr)^2$ exhausts all the square divisors of $mn$ when $q^2$ and $r^2$ run over square divisors of $m$ and $n$ respectively, verifying the last equality in eqn.\eqref{Jm_Jn}.
Since the Hecke image is a sum over representatives in the right coset $ \Delta^{(n)}_N=\Gamma(N)\backslash [\![ J_n ]\!]$, the multiplication rule eqn.\eqref{Tm_Tn:new} follows as a result of eqn.\eqref{Jm_Jn}.

To prove eqn.\eqref{T_p^n:new}, we only need to verify the relation among the sets of representatives. Recall the representative associated with the prime power $p^n$ 
\begin{align}\label{representative:p^n}
\Delta_N^{(p^n)}=
\bigcup_{j=0}^n
\left\{ 
\sigma_{p^{n-j}}
\begin{pmatrix}
p^{n-j} & \lambda_j N \\0 & p^j
\end{pmatrix}
\Bigg|
0\leq \lambda_j < p^j
\right\}.
\end{align}
One can show that
\begin{align}\begin{split}
& \Delta_N^{(p)} \circ 
\left\{ 
\sigma_{p^{n-j}}
\begin{pmatrix}
p^{n-j} & \lambda_j N \\0 & p^j
\end{pmatrix}
\Bigg|
0\leq \lambda_j < p^j
\right\} 
\\
=& \sigma_{p^{n-j}}\circ \Delta_N^{(p)} \circ 
\left\{ 
\begin{pmatrix}
p^{n-j} & \lambda_j N \\0 & p^j
\end{pmatrix}
\Bigg|
0\leq \lambda_j < p^j
\right\} \\
=& \left\{   \sigma_{p^{n-j}} 
\begin{pmatrix}
1 & (\lambda_j+p^j\nu) N \\0 & p^{j+1}
\end{pmatrix}
\Bigg| 
0 \leq \lambda_j < p^j  ,~ 0  \leq \nu < p
\right\} 
\\
&+
\begin{pmatrix}
p & 0 \\ 0 & p
\end{pmatrix}
\sigma_p
\circ 
\left\{ 
\sigma_{p^{n-j}}
\begin{pmatrix}
p^{n-j} & \lambda_j N \\0 & p^{j-1}
\end{pmatrix}
\Bigg|
0\leq \lambda_j < p^j
\right\} 
\end{split}\end{align}
Summing over $j$ from $0$ to $n$,
one establishes the relation among double cosets
\begin{align}
[\![ J_p ]\!] [\![ J_{p^n} ]\!] 
= [\![ J_{p^{n+1}} ]\!] 
+ p (p \, \sigma_p) [\![ J_{p^{n-1}} ]\!] ,
\end{align}
where the factor $p$ in front of $(p\,\sigma_p)$ means $p$ copies. The $p\,\sigma_p$ in the parentheses is understood as $p\mathbb{I}_2 \cdot \sigma_p$, a composition of the homothety operator $p\mathbb{I}_2$ and the action $\sigma_p$.
The homothety operator $p\mathbb{I}_2$ is a rescaling of the lattice and acts trivially on $\tau$.
However, it induces a factor $p^{-k}$ via the factor $(c\tau+d)^{-k}$ upon a $GL(2,\IZ)$ action, completing the proof of eqn.\eqref{T_p^n:new}.

\end{document}